\documentclass{pas}
\usepackage{lineno}
\usepackage{aas-macros}
\usepackage{multirow}

\begin{document}
\lefttitle{\texttt{Frabjous}: Deep Learning Fast Radio Burst Morphologies}
\righttitle{Ajay Kumar et al.}

\jnlPage{13}{7}
\jnlDoiYr{2025}
\doival{10.1017/pasa.xxxx.xx}

\articletitt{Research Paper}

\title{ \texttt{Frabjous}: Deep Learning Fast Radio Burst Morphologies}
\author{\sn{Ajay} \gn{Kumar}$^{1}$, \sn{Ashish} \gn{A.~Mahabal}$^{2,3,4}$ and \sn{Shriharsh} \gn{P.~Tendulkar}$^{1,5,6}$}

\affil{$^1$National Centre for Radio Astrophysics, Post Bag 3, Ganeshkhind, Pune, 411007, India \\
$^2$Division of Physics, Mathematics and Astronomy, California Institute of Technology, Pasadena, CA 91125, USA \\
$^3$Center for Data Driven Discovery, California Institute of Technology, Pasadena, CA 91125, USA \\
$^4$Inter-University Centre for Astronomy and Astrophysics, Pune, 411 007, India \\
$^5$Department of Astronomy and Astrophysics, Tata Institute of Fundamental Research, Mumbai, 400005, India \\
$^6$CIFAR Azrieli Global Scholars Program, CIFAR, Toronto, Canada}

\corresp{A. Kumar, Email: akumar@ncra.tifr.res.in}



\begin{abstract}
The increasing field of view of radio telescopes and improved data processing capabilities have led to a surge in the detection of Fast Radio Bursts (FRBs).
The discovery rate of FRBs is already a few per day and is expected to increase rapidly with new surveys coming online.
The growing number of events necessitates prioritized follow-up due to limited multi-wavelength resources, requiring rapid and automated classification.
In this study, we introduce \texttt{Frabjous}, a deep learning framework for an automated morphology classifier with an aim towards enabling the prompt follow-up of anomalous and intriguing FRBs, and a comprehensive statistical analysis of FRB morphologies.
Deep learning models require a large training set of each FRB archetype, however, publicly available data lacks sufficient samples for most FRB types.
In this paper, we build a simulation framework for generating realistic examples of FRBs and train a network based on a combination of simulated and real data starting with the CHIME/FRB catalog.
Applying our framework to the first CHIME/FRB catalog, we achieve an overall classification accuracy of approximately 55\%, well over a random multiclass classification rate of 20\% with five balanced classes during training. While this falls short of desirable performance, we critically discuss the limitations of our approach and propose potential avenues for improvement. Future work should explore strategies to augment training datasets and broaden the scope of FRB morphological studies, aiming for more accurate and reliable classification results.
\end{abstract}


\maketitle

\section{Introduction} \label{sec:intro}

Since the discovery of the first Fast Radio Burst (FRB) in archival data of Parkes radio telescope \citep{lorimer2007}, we have more than thousand FRB discoveries published \citep{2022A&ARv..30....2P} and few thousands more FRBs that have been discovered \citep{scholz2022_aas}. The mechanisms of FRB emission and the astrophysical channels through which they are formed are not yet understood. There are multiple theoretical models discussing the origins of FRBs \citep{platts2019} like from magnetar \citep{beloborodov2020blast,kulkarni2024frbfrommagnetars,murase2016_frb_modelmagnetar,katz2016soft,Metzger2019,lyubarsky2014model}, NS-NS merger \citep{Totani2013,Wang2016_frbfromnsnsmerger}, NS-BH merger \citep{Mingarelli_2015}, etc. However, yet no one specific model has been unambiguously supported by observations. Some FRBs repeat while other FRBs have been followed up for hundreds of hours with no detectable repetition. It is not clear whether the repeating and non-repeating FRBs originate from the same astrophysical channels or whether they represent distinct origins. Understanding and identifying the origins of FRBs and their relation to other transients requires a multi-pronged approach of multi-wavelength follow-up.

One aspect of understanding FRB origins is to study their hosts and the local environments in which they form. The arcsecond and sub-arcsecond localisation of FRBs using intermediate and very long baseline interferometry \citep[][]{2017ApJ...834L...8M,2017Natur.541...58C}, has allowed the study of FRB environments in dwarf galaxies \citep{2017ApJ...843L...8B,2017ApJ...834L...7T}, in large early-type galaxies \citep{2020ApJ...895L..37B}, in regions of active star-formation \citep{2021ApJ...908L..12T} and also in globular cluster locations with a very old stellar population \citep{2022Natur.602..585K}. The spatial resolution is required to study the immediate environments of source or origin which would be crucial to study the nature of FRB progenitors. CHIME/FRB collaboration has used VLBI localization to study correlations between FRB 20210603A and the disk of its host galaxy \citep{2023arXiv230709502C}. The increased spatial resolution with VLBI  allows us to study the source and environment of FRBs. 
\par
Another approach to studying the origins of FRBs is to identify their links to other transients and undertake multi-wavelength, multi-messenger follow-up for their prompt counterparts. To date, SGR 1935+2154 is the only known source that has emitted two $\sim$ millisecond long radio bursts along with near-simultaneous hard X-ray bursts \citep[][]{2020ApJ...898L..29M,2020Natur.587...59B,2020ATel13681....1S}. No extragalactic FRB has a well-identified multi-wavelength or multi-messenger counterpart, though many upper limits have been placed on the X-ray or $\gamma$-ray \citep[][]{2020ApJ...888...40A,2021ApJ...915..102V,2021NewAR..9201595F}{}{}, optical/near infrared \citep[][]{2023arXiv231109316K, 2021ApJ...907L...3K}, gravitational waves \citep{Ligo-chime-2022}, and neutrino emissions linked with FRBs. There has been one suggested association of FRB 20190425A and a binary neutron star merger GW190425 that happened about 2.5 hours before the FRB \citep{2023NatAs...7..579M}. However, \citep{Smartt2024} disfavour that the association but do not disprove it based on ATLAS and Pan-STARSS survey.
\par
However, the rapid follow up of FRBs is challenging due to their large sky rate \citep[$800\,\mathrm{sky^{-1}\,day^{-1}}$ at a fluence $>5$\,Jy-ms at 600 MHz][]{2021ApJS..257...59C, 2023ApJ...947...83C}, far higher than the observed rates of any other short transients such as gamma-ray bursts, magnetar flares, and compact binary coalescences. Most FRBs are not expected to have detectable prompt counterparts, making it impractical to followup every detection. Furthermore, the prompt followup resrources are scarce. Due to high sky rate and limited follow-up resources  we require to identify intersting cases for multi-wavelength and possibly multi-messenger followups to uncover the FRB origins. Any detection would be tranformative. Several efforts have been made to rapidly follow up FRBs to search for their prompt counterparts, afterglows, and multi-messenger counterparts. The Gamma-Ray Urgent Archiver for Novel Opportunities (GUANO) system and rapid follow up on the \emph{Swift} mission \citep{tohuvavohu2020} has enabled  rapid follow up of FRBs as well as gravitational wave mergers \citep{oates2021}. However, resources for prompt FRB follow-up are scarce and the increasing rates of FRB detections from CHIME, ASKAP, DSA-110, MeerKAT make it essential to prioritize follow up to FRBs that are rare and anomalous in some aspect. The reduce timescales of the follow-up efforts make it necessary to create an automated prioritisation framework for FRB follow-up. For example, \citet{lin2023a, lin2023b} identified ultra-bright and rare FRBs detected in the far sidelobes of the CHIME/FRB telescope based on the spectral signatures of diffraction and used them to constrain the rate of repetition as well as the local dispersion measure. However, these identifications were done well after the detection of the FRBs, and with human inspection. Similarly, recently discovered FRB\,20250316A, a ultra bright nearby FRB (S/N $\geq$ 5000) demonstrates that nearby ultra-bright FRBs can be detected in the side lobe of CHIME with modest S/N \cite{2025ATel17081....1N}. These events are ultra-bright due to their closer proximity and are most promising in getting clues about their origins from multi-wavelength follow-up. Recently, \citep{Hanmer_2025} conducted the first ever near-simultaneous optical observations for a non-repeating FRB. They used to MeerLICHT to observe just $\sim$ 3.4s after FRB 20230808F was detected by MeerKAT telescope, providing constraints on FRB progenitor models.

Many repeating FRBs have shown distinct morphological structure in their dynamic spectra (i.e. `waterfall' plots). These include the well-known downward drifting subcomponents \citep[i.e. the `sad trombone'; ][]{2014ApJ...790..101S, 2019ApJ...876L..23H}, as well as narrow-bandedness, and multiple components. \citet{2021ApJ...923....1P} presented a detailed study of morphology of the 535 FRBs from the first CHIME/FRB catalog \citep{2021ApJS..257...59C}. They identified four major archetypes and reported on average repeating FRBs have temporally wider and spectrally narrower bursts than non-repeating FRBs. Further, rare FRBs with atypical structures such as  sub-millisecond quasi-periodicity \citep{2023A&A...678A.149P}, ultra bright nearby FRB \citep{2025ATel17081....1N} that could be detected in a sidelobe and a variety of morphological structures \citep{2022NatAs...6..828C}.   

We aim to build a predictive classifier that can identify different morphological types of FRBs, in the futrue highlight anomalous FRBs, and extract the structural parameters for the bursts. Most previous efforts in classifying fast transients have focused on distinguishing FRBs from radio frequency interference (RFI) which is critical when searching for FRBs in real-time. \citet{2018AJ....156..256C} for the first time used deep learning framework for single pulse classification. They utilized dynamic spectra, time series, and multibeam signal to noise ratio (SNR) as inputs to several independent deep neural networks (DNNs) which can classify a candidate as FRB or RFI. \texttt{FETCH}\footnote{\url{https://github.com/devanshkv/fetch}} \citep{2020MNRAS.497.1661A, 2020ascl.soft05014A} is a widely-used deep learning based binary classifier for FRB/RFI separation. \texttt{FETCH} leverages transfer learning by using initial convolution layers pre-trained on the \texttt{ImageNet} dataset and is very good at feature extraction. They use simulated and real FRB and RFI data to train the classifier where weights are only updated for the densely connected layers at the end of network architecture. \cite{Yadav_2020} developed \texttt{IntensityML} using deep learning to develop models that can classify between FRB and RFI for CHIME/FRB backend. 
\par
Others have used clustering and unsupervised learning techniques to identify different classes of FRBs. \citet{2023MNRAS.518.1629L} studied the differences between repeating and non-repeating FRBs based on the parameters such as fluence, box-car width, energy, and excess DM. 
\citet{2023MNRAS.519.1823Z} applied various clustering algorithms to the CHIME/FRB catalog to try to understand if some non-repeaters are appearing to be so due to lack of observations and whether there are clear differences between the parameter spaces for repeaters and non-repeaters. However, these efforts are limited by the small and imbalanced dataset for repeaters and non repeaters. Machine learning algorithms such as UMAP and t-SNE have been used to classify long and short GRBs \citep{2023ApJ...945...67S}, to classify repeating and non-repeating FRBs \citep{2022MNRAS.509.1227C,2023MNRAS.522.4342Y} or extremely repeating FRBs \citep{2023MNRAS.521.5738C}. Recently, \citet{kuiper2025representationlearningfastradio} showed that FRBs may not be naturally clustered in parameter space, and that techniques such as representation learning will be required to identify distinct classes. A larger, more comprehensive sample of FRBs is crucial for machine-learning algorithms to classify them robustly \citep{2023MNRAS.522.4342Y}. 
\par
In this paper, we present \texttt{Frabjous}, a framework for simulating various different FRB morphologies and training a deep learning classifier to identify FRBs based on their dynamic spectra. This is intended for 
\begin{enumerate}
    \item a  well-characterised understanding of the statistics of FRB morphologies,
    \item identification of possibly anomalous FRBs, and
    \item improved prioritisation of rapid FRB follow-up.
\end{enumerate}
The archetypes and models used in this paper are necessarily preliminary and are expected to be updated as we learn more about different types of FRBs. These are intended as a starting point for a broader classification effort, where anomalous FRBs correspond to events that do not conform to any of the identified distinct morphological classes in this work and therefore lie outside the scope of the present classification scheme. The paper is organised as follows: in Section~\ref{sec:simulation}, we describe our simulation framework and prescriptions/recipes for simulating a variety of observed FRB morphologies; in Section~\ref{sec:classfication}, we describe and train two different sets of deep learning models --- a combination of binary classifiers and a multi-class classifier; in Section~\ref{sec:training}, we discuss the challenges for the application of the classifier to real data from CHIME and the performance of the classifier with the real data i.e. we take the CHIME/FRB first catalog in Section~\ref{sec:testing_with_real data}. Finally, in Section~\ref{sec:discussion} , we discuss the results of our classification, limits of applicability, and future extensions of this framework.

\section{Simulation of FRB Morphologies}   \label{sec:simulation}

Within the $\approx 1000$ FRBs that have been published from various surveys, there is a wide variety of morphologies, particularly between repeaters and non-repeaters \citep{2021ApJ...923....1P}. However, we cannot directly use just these FRBs as a training set since they are few, observed by different telescopes, and the different morphologies are not equally represented. Such imbalanced, heterogenous datasets can cause biases in the classifier. In addition, each FRB search pipeline has its own set of biases \citep[see e.g. ][]{2023AJ....165..152M} which can skew the distribution of burst population parameters. Using simulated bursts, several combinations of burst features which are not generally detected or seen -- but are not unexpected -- can be incorporated into the training set (e.g. wide scattered bursts or wide narrow-band bursts). Hence, our model can be trained on types of bursts not commonly detected. This will not only make the classifier telescope-agnostic, but also ready for use by upcoming telescopes with greater sensitivity over a bigger parameter space. 

\subsection{Simulating training data}
The input to the classifier is an 2-D array representing the nearly de-dispersed intensity dynamic spectrum of each FRB. To simulate these dynamic spectra for FRBs, we utilize the \texttt{simpulse}\footnote{\url{https://github.com/kmsmith137/simpulse}} library, which is capable of generating a dispersed single pulse in a time series across a given frequency range with a specified number of frequency channels. We write a wrapper around \texttt{simpulse} to compose FRBs with arbitrary numbers of components and spectral structures. In this section, we describe the methods and parameters used to simulate different FRB morphology.

\begin{figure*}
  \centering
  \includegraphics[width=\textwidth]{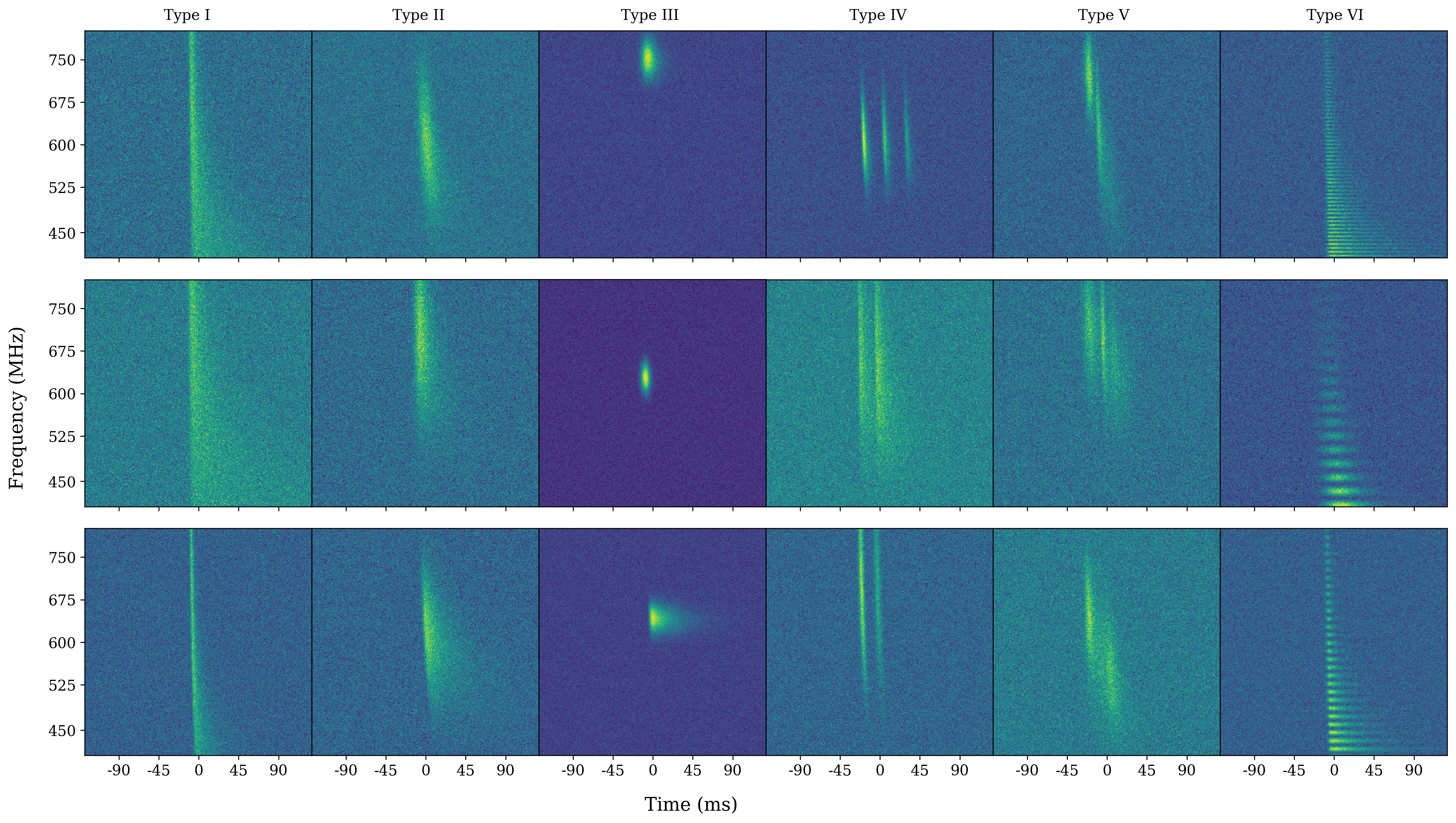}
  \caption{ In each row from left to right example of type I, II, III, IV, V and VI burst morphology simulated from our framework. We have randomly chosen three bursts from simulations for each type to demonstrate the different types.}
  \label{fig:example_bursts}
\end{figure*}

We expand upon the various burst morphologies identified in \citet{2021ApJ...923....1P} using the first CHIME/FRB catalog bursts. The burst morphologies we use in this paper are not physically motivated but the visual difference in the structure of the detected emission in the dynamic spectra, e.g. see Figure 3  of \citet{2021ApJ...923....1P} and also illustrated in Figure~\ref{fig:example_bursts}. For all the simulated bursts, we have used a fiducial observation bandwidth of $400-800$ MHz and a sampling time of 1\,ms.  In our simulations, we introduce a small random dispersion value to account for the error dispersion measure (DM) estimates during detection. According to \citet{2021ApJ...923....1P}, there is a reported bias in the estimation of DM, with an overestimation ranging from $0.5$ to $1\,\text{pc cm}^{-3}$ for the first CHIME/FRB catalog.

The burst is generated using a Gaussian temporal profile for the pulse, and the intrinsic width of the pulse is defined as the Full Width at Half Maximum (FWHM) of the Gaussian profile. \texttt{simpulse} introduces scattering across the observing bandwidth. The spectrum of each burst component is defined by a Gaussian profile, a power-law with spectral running \citep[as defined in ][]{2021ApJ...923....1P}, or a diffraction pattern to imitate the far sidelobe FRBs \citep[][]{lin2023a}. 

For this paper, we choose six categories and for each few simulated examples : 

\begin{enumerate}
    \item Type I: Single component, broadband power-law-like spectrum spanning the observation bandwidth. 
    \item Type II: Single component, Gaussian-like spectrum partially covering the observational bandwidth. The FWHM of the Gaussian varies from 100 MHz to 400 MHz with the central frequency ranging from one edge of the band to another. 
    \item Type III: Single component, with Gaussian-like spectrum with a very small bandwidth the fractional bandwidth is $<25$\%.  
    \item Type IV: Multi component burst with each having a similar spectrum, derived from range of spectra seen for first CHIME/FRB Catalog bursts.
    \item Type V: Multiple bursts showing a downward-drifting pattern i.e. central frequency of each component decreases with time. Each component has a gaussian spectrum.
    \item Type VI: This is inspired by \citet{lin2023a}. Sharp multiple peaks along the frequency axis are observed because of the diffraction pattern caused by a detection in the far sidelobe of the telescope. 

\end{enumerate}


\begin{table}
\footnotesize
\caption{Parameter ranges for different FRB morphologies. For multiple components, these ranges define each sub-burst. All parameters are sampled from uniform distributions, except for the width, which combines a log-normal distribution and a uniform distribution for widths $\geq 10$ ms.}
\label{tab:Table_1}
\centering
{\tablefont
\setlength{\tabcolsep}{3pt}
\begin{tabular}{lccccc}
\toprule
\textbf{Type} & \textbf{Width } & \textbf{Spectral Index$^{\rm a}$} & \textbf{Bandwidth } & \textbf{Central Frequency } \\
\textbf{} & \textbf{(ms)} & \textbf{} & \textbf{ (MHz)} & \textbf{(MHz)} \\
\hline
I   & 1 -- 30   & --3 -- +3 & 400         & 600             \\
II  & 1 -- 30   & ---        & 100 -- 400  & 450 -- 750      \\
III & 1 -- 30   & ---        & 10 -- 50    & 425 -- 775      \\
IV$^{\rm b}$ & 1 -- 15   & ---        & ---         & ---             \\
V   & 1 -- 10   & ---        & 50 -- 100   & 425 -- 775      \\
VI  & 1 -- 30   & --3 -- +3 & 400         & 600             \\
\botrule
\end{tabular}}
\begin{tabnote}
{$^{\rm a}$Spectral index is only defined for Types I and VI.}\tnp
{$^{\rm b}$Spectral features for Type IV are described in the text.}\tnp
\end{tabnote}
\end{table}

For each type, we describe different features and ranges of burst parameters are listed in Table~\ref{tab:Table_1}. We did not have sufficient numbers to construct reliable distributions for most parameters. Hence, we take uniform distributions with bounds provided by the typical range observed in FRBs, particularly in the first CHIME/FRB catalog. The following points describe the methods to generate each type of FRB morphology: 

\paragraph{Type I:} We simulate type I bursts using \texttt{simpulse} with the range of the parameters as in Table~\ref{tab:Table_1}. We introduce power law-like spectra on the broadband pulse as seen in galactic pulsars with a spectral index sampled uniformly in the range $[-3,+3]$. The widths of the bursts are sampled as described at the end of this subsection.

\paragraph{Type II:} All the features except the spectral shape are similar to type I bursts. The narrowband emission is modelled by Gaussian profile instead of a power law.

\paragraph{Type III:} This is similar to the types I and II except the visible bandwidth of the emission is much smaller.   

\paragraph{Type IV:} We have used the values of spectral index, spectral running from the CHIME/FRB Catalog. To generate the multi-component structure, we use the recipe below:


\begin{enumerate} 
   \item Choose the number of sub-bursts: Randomly and uniformly chosen to be 2 or 3. 
   \item Choose the arrival time separations between components: Uniformly distributed between 5\,ms to 25\,ms. 
   \item For each sub-burst, we set the fluence for other components with a ratio between 0.2 and 1, compared to the first component's fluence. 
   \item Pulse width ratio of the second (and of the third, when present)  components is between 0.8 to 1.2 to the first component.
   \item We vary the  scattering and DM error similar to the types mentioned above. 
\end{enumerate}   

\paragraph{Type V:} Here the individual sub-bursts are simulated similar to type IV, but we use the following procedure to simulate the downward drifting sub-structure: 

\begin{enumerate}
    \item We randomly choose drift rates from a uniform distribution between $[-1, 20]\,\mathrm{MHz\,ms^{-1}}$, where the negative downward drift rate represents a rare upward drifting structure. We agnostically assume a uniform distribution for drift rates. 
    \item The number of components is randomly chosen from a uniform distribution between 2 to 5. The arrival time separations and fluxes are chosen as for type IV.
    \item The central frequency of the brightest component is chosen between [425, 725]\,MHz.
    \item With the choice of drift rate and arrival time, the central frequencies of the other components are chosen to lie along the drift rate.
    \item Each component's bandwidth is randomly chosen from a uniform distribution ranging between [50, 100]\,MHz.
    
\end{enumerate}



We include the prescription for type VI in appendix~\ref{sec:typeVI_prescription} as they are not discussed in the rest of the paper. In Figure~\ref{fig:example_bursts}  we show some of the simulated bursts for each type. Typically, we generated 1000 samples for each type.

\begin{figure}{}
 \includegraphics[width=0.45\columnwidth,trim= 2.7cm 1.3cm 4cm 1cm, clip]{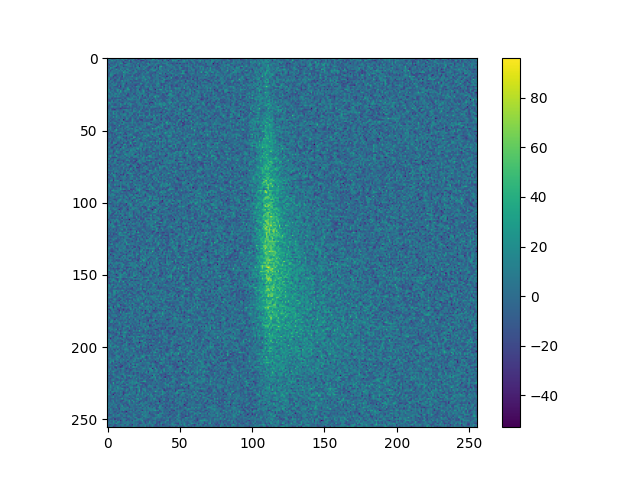}
 \hfill
 \includegraphics[width=0.45\columnwidth,trim= 2.7cm 1.3cm 4cm 1cm, clip]{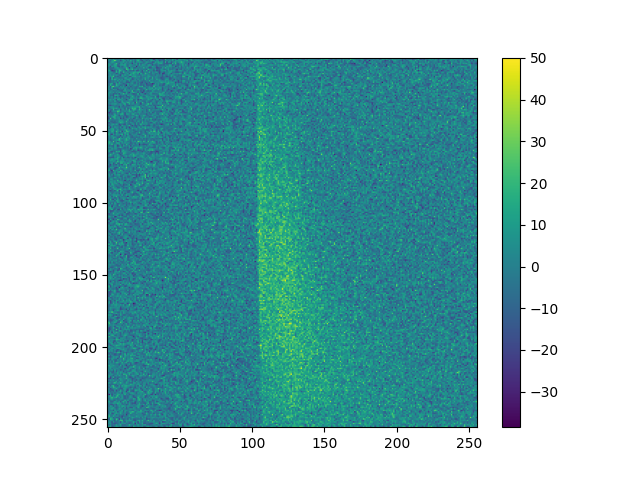}
 \vfill
 \includegraphics[width=0.45\columnwidth,trim= 3.6cm 1.3cm 3cm 2cm, clip]{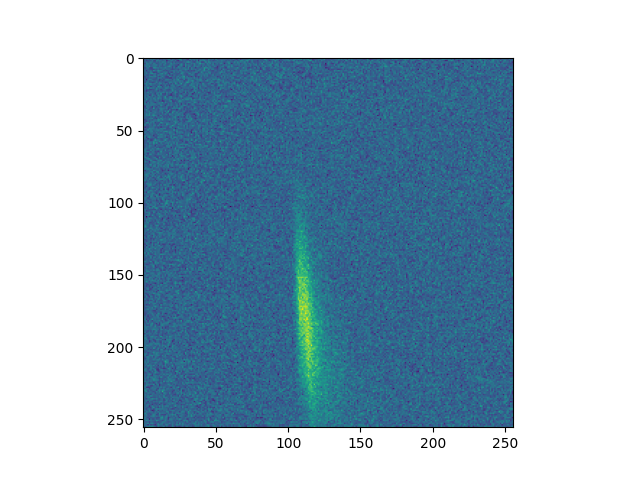}
 \hfill
 \includegraphics[width=0.45\columnwidth,trim= 3.6cm 1.3cm 3cm 2cm, clip]{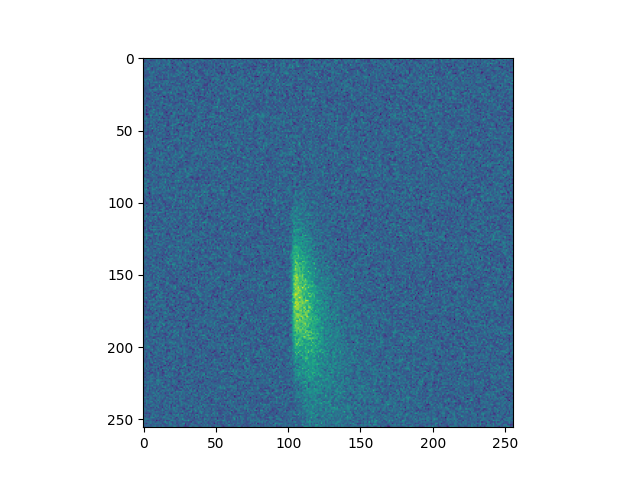}
 \caption{Examples of type IV (top row) and of type V (bottom row) where there is overlap between successive components and can be easily confused for single component bursts (e.g. type II).}
 \label{fig:bad_C1_types}
\end{figure}

\paragraph{Width distributions}
We sample widths for 90\% of the FRBs from a log-normal distribution with parameters from the CHIME/FRB catalog. The remaining 10\% of the widths are sampled from a uniform distribution to incorporate wider bursts. For types I, II and III, we take a uniform distribution from 10 ms to 30ms. For type IV and type V we take the range from 10 ms to 15 ms to avoid overlap of successive components. There were type IV and type V bursts that still have some overlap between the successive components and it becomes difficult with the eye to distinguish them. To make the separation between successive components clear we constrain the arrival time separation to be greater than half the pulse width of the following components. Despite this constraint, some bursts at low signal to noise ratio do not show visibly distinct components. Few of these examples are shown in Figure~\ref{fig:bad_C1_types} for type IV and type V respectively. We exclude such samples from training data, at millisecond resolution, they visually appear as single-component events even to the eye. Including them would bias the training set, since the model would learn noise or resolution driven artefacts rather than true morphological structure. In real datasets, e.g., \citep{curtin2025morphology35repeatingfast}, some Type-IV bursts are only revealed at microsecond resolution; at millisecond sampling, they also look simpler. We filter those samples manually.

\subsubsection{Signal to noise ratio \& noise distribution}

We simulate FRBs with a wide range of signal to noise ratios (SNRs; as defined below) to replicate the SNR distribution of detected bursts. We choose SNRs of 15, 25, 35, 50, and 100. We do not use bursts fainter than SNR 15 since the morphology is not very visible, even for human inspection. The underlying noise distribution is chosen to be Gaussian for simplicity. We discuss the implications of this choice below and note that the FRB simulation can be easily added into real noise from a telescope or a more sophisticated realisation of noise.

In the definition of SNR, the width of the pulse is typically defined by the FWHM of the Gaussian profile fitted to the data. However, this definition does not work for multi-component bursts. To maintain uniformity in determining the width of the bursts, we use the $T_{90}$ width  as is generally done in GRB measurements first described in \citep[][]{1996PhDT.........7K}. This is the time between the cumulative burst profile rising from 5\% to 95\% of the total emission, encompassing $90\%$ of the fluence. We apply this method to determine widths for all types of bursts, i.e. single and multi-component bursts. A Table is provided in the Appendix~\ref{sec:snrcomparison} for some of the bursts from the first CHIME/FRB catalog to give an idea of how the SNRs compare with typical boxcar width estimation and T90 definition. 

After determining the width for each type of FRBs, we calculate the SNR using Equation~\ref{eq:equation1}, where $\sigma$ is the root-mean-square noise (per time sample) in the band-averaged time profile, $F$ is the band-averaged fluence, $w$ is the $T_{90}$ width calculated as above, and $S$ is the estimated SNR.
\begin{equation}
 S = \frac{F}{\sigma \sqrt{w}}
     \label{eq:equation1}
\end{equation}


\subsection{Data Augmentation}

Data augmentation like flipping and rotation are not applicable in our case because several desired features will be lost. These simulated FRBs have features like scattering, and downward drifting along one direction in time. Any such data augmentation would lead to changes in the original structure intended for the particular type. However, for training, we can simulate an adequate number of FRBs so that each binary classifier can learn to distinguish between two different classes. 

\subsection{Caveats of the simulation framework}
\label{sec:limitations of simulations}
While we have addressed most of the features currently observed in published FRBs, there are still some limitations to the simulated training dataset which we discuss here. We simulated the dataset by considering an observing frequency of 400 to 800 MHz, reflecting the current focus on CHIME telescope data in most published FRBs. However, other telescopes such as ASKAP, DSA-110, MeerKAT, FAST are operating at different frequencies, channelization, and time resolutions. This has implications for downward drifting sub-pulses, scattering time, and resolvability of multi-component bursts. For each telescope, the training set can be suitably modified to retrain the classifier.

Our simulations cover timescales in the milliseconds, as commonly seen for FRBs. However, it is worth noting that some bursts have been observed over timescales of microseconds and that seemingly single component bursts at millisecond timescales show complex morphologies at microsecond timescales  \citep[e.g.][]{2022NatAs...6..393N, 2023MNRAS.526.2039H, Sand_2025}. 
The drift rate range used for simulating type V is broad, ranging from -1 $\,\mathrm{MHz\,ms^{-1}}$ to -60$\,\mathrm{MHz\,ms^{-1}}$, but in our simulation, it is limited to -20$\,\mathrm{MHz\,ms^{-1}}$ to ensure that the sub-bursts lie within the observing bandwidth. Additionally, at higher frequencies up to a few GHz with larger bandwidth, drift rates lower than -60$\,\mathrm{MHz\,ms^{-1}}$ can occur.
\par
In our simulations, we only added Gaussian noise to the dynamic spectra. A more robust method would involve simulating bursts in the presence of different noise types, such as complex noise that mimics telescope noise. This is quite important for the classifier's performance under realistic conditions. Adding only Gaussian noise to the simulated FRBs makes the classifier telescope-agnostic, albeit at the cost of sub-optimal performance. However, the model can be retrained with telescope-specific complex noise to achieve optimal performance for a particular telescope backend. We also note that most FRBs are detected in `quiet' periods with little to zero bursty RFI, so the assumption of a Gaussian noise distribution may not too sub-optimal. Secondly, most telescope pipelines handle the more persistent narrow-band RFI by masking the offending channels. In Section~\ref{sec:testing_with_real data}, we show how we can interpolate between masked channels to use our framework on real data.
\par
Lastly, while we assumed a uniform distribution for all burst parameters except for widths, it is important to note that the current population may not be sufficient to determine any distribution in burst parameters and that distribution of measured parameters like fluence and width can be biased due to telescope design and detection algorithms.

\subsection[short]{Dataset preparation for training}

Using python scripts we initially generated 1000 samples for each SNR values of 100, 50, 35, 25 and 15 for each FRB type. Subsequently, this data was split into training, validation, and test sets. Each simulated burst is essentially a 2-D \texttt{numpy} array measuring $256\times256$ pixels in frequency (400--800\,MHz) and time (256\,ms). The $256\times256$ image size allows us to capture all the relevant features and accommodate samples with wider widths. We aimed to keep the input array size minimal without compromising the features needed for the convolution neural network (CNN) layers to extract local pattern in the images.

\section{Framework for Classification  }
\label{sec:classfication}

In this section, we describe the basic architecture of neural networks based on deep learning for the binary classification. Each binary classifier is used to distinguish between a pair of FRB classes. \\
We also construct a single multi-class classifier that is used to classify the first five categories as shown in Figure~\ref{fig:example_bursts}.

\subsection{Network Architecture}
\label{sec:Network architecture}
We use \texttt{keras} \citep{chollet2015keras} and \texttt{tensorflow} \citep{abadi2016tensorflow}  for the model framework. 
We follow standard methods to develop a deep learning framework for image classification where the first few layers are CNN layers to extract local features in the images. Filters or kernels in each layer extract specific local patterns in the images and then  densely connected layers extract the global features in the images. Each layer contains multiple filters with  \texttt{relu} for activation functions. 

\begin{figure*}{}
  \centering
  \includegraphics[width=\textwidth,trim= 0.5cm 14.8cm 0.3cm 0.65cm, clip]{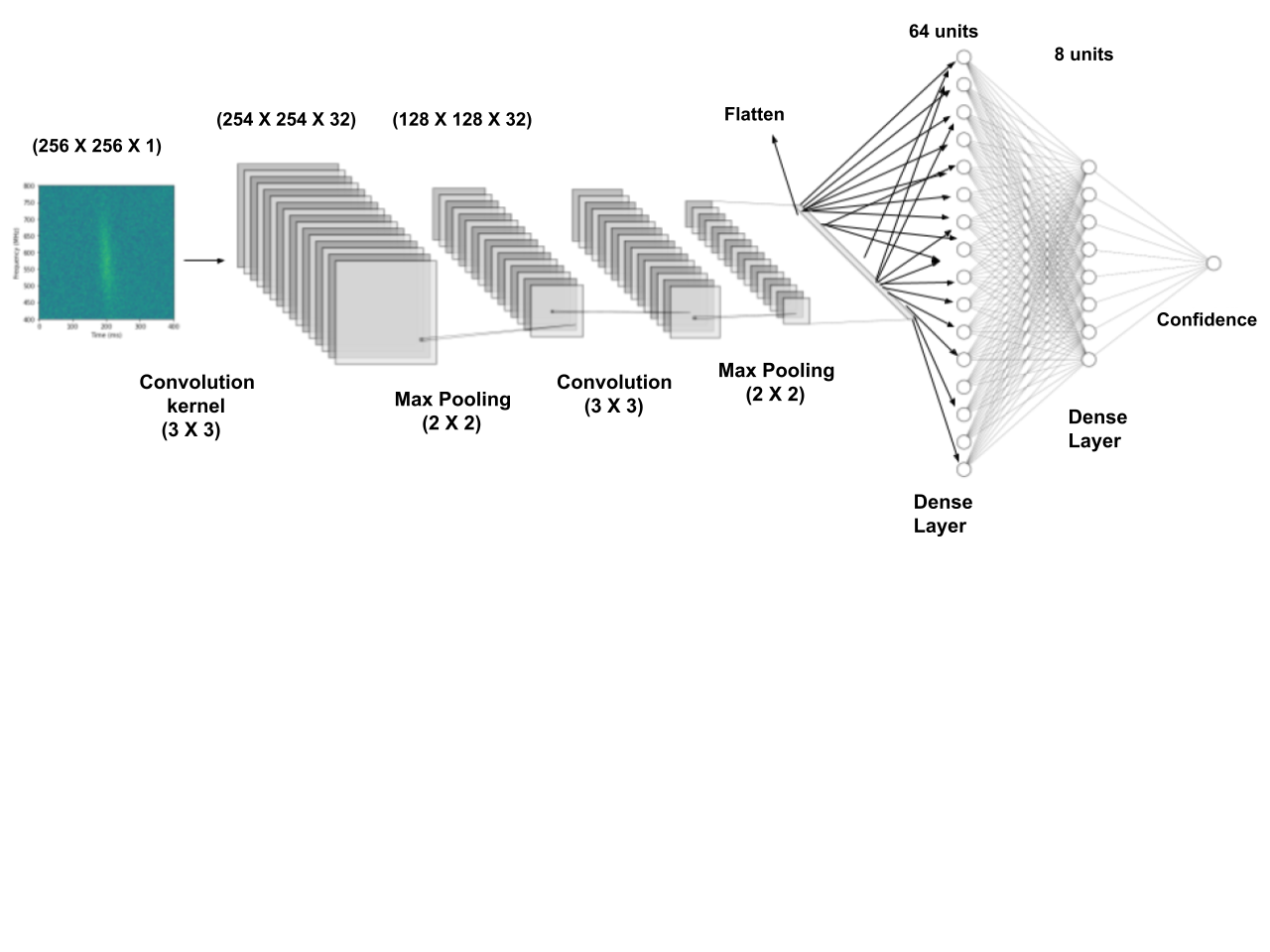}
  \caption{A schematic diagram of a typical binary classifier network. A dynamic spectrum ($256\times256$) is the input for the classifier. The first few layers are convolutional, the next layers are fully connected layers, and the final layer is one that gives the confidence of the input belonging to the classes under consideration. Figure made using \texttt{NN-SVG} \citep{LeNail2019}.  }
  \label{fig:architecture}
\end{figure*}

\begin{figure}
  \centering
  \includegraphics[width=\columnwidth,trim= 0.5cm 2.2cm 10.0cm 2.0cm, clip]{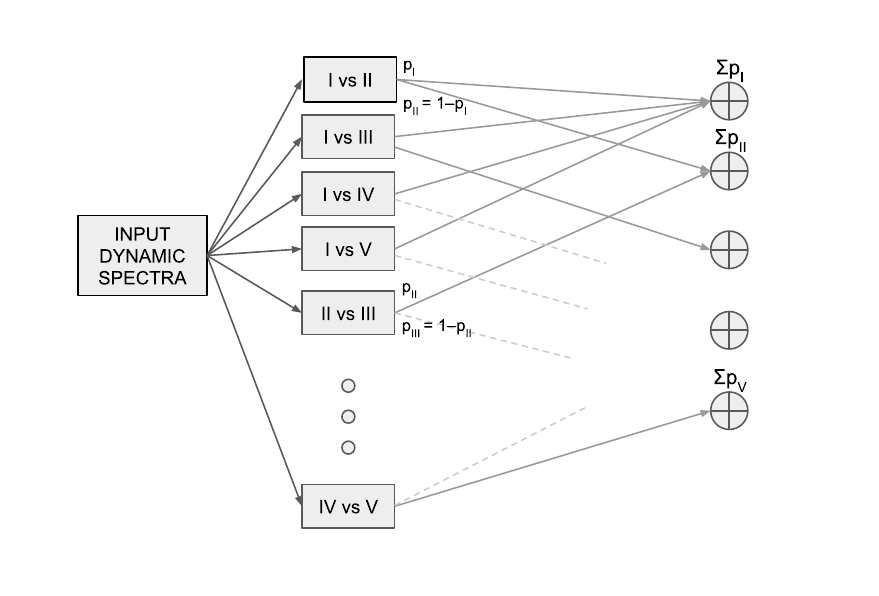}
  \caption{ A schematic diagram illustrating the workflow where input dynamic spectra is processed through multiple binary classifiers (with the first class labeled as negative and the second as positive). The outputs from these binary classifiers are then combined to infer the final classification, detailed in Section~\ref{sec:multiclass_with_binarymodels}. For clarity, only a subset of binary classifiers is shown, and dashed and solid lines indicate how outputs from individual classifiers are mapped to their respective types.}
  \label{fig:illustration_one_vs_one}
\end{figure}

We used \texttt{Adam} \citep{kingma2017adam} optimiser which is known to perform best for binary image classification tasks using the deep learning models. We use dropout layers to avoid overfitting.  
A \texttt{sigmoid} activation function is used in the last dense layer which outputs the confidence for input image belonging to the positive class. Figure~\ref{fig:architecture} illustrates a typical network architecture we use for distinguishing between two FRB types. Similar architectures --- with slight variations in the parameters for the model framework --- are  used to distinguish between other pairs. We develop 10 (${=}^5{C}_2$) binary classifiers accounting for all pairs of classes and combine their output to build a multi-class classifier. We pass the input dynamic spectrum through each of the 10 binary classifiers and sum up the output confidence for each class (Figure~\ref{fig:illustration_one_vs_one}). The output confidence is ideally scaled in the range from -0.5 to 0.5, where 0.5 indicates high confidence identification of one class and -0.5 indicates the other class. In the final implementation, we tweaked the distinction threshold slightly based on each classifier's false positive and false negative rates. The final output is a combined score of confidence of each type. The details of this combination of scores are discussed in Sections~\ref{sec:multiclass_with_binarymodels}. We also use the single multi-class classifier with similar as shown in Figure~\ref{fig:architecture} but a larger network to distinguish all five classes at once. In the end, we use \texttt{Softmax} activation function for the last dense layer, giving confidence for each class as output with maximum for the positive class.


\section{Training }
\label{sec:training}

In this section, we describe our training process. 
For initial experiments of training binary classifiers, we divide the dataset of 2000 samples including all SNR values into 75\% training, 25\% validation. For testing we take 500 samples for each SNR value. 

We know that a trained model has to learn the features in presence of different levels of noise typical to FRBs. Hence we need to use a mixture of different SNRs for training. To confirm this we did some initial experiments by training with samples having identical SNR. We employed a dataset with samples having SNR of 100 to train a base binary classification network for each pair of classes. To understand  this dependence of performance with SNR, subsequent experiments involved training the model on lower SNR datasets, specifically SNR 15. While the models exhibited significantly improved performance at SNR 15, its performance suffered when confronted with higher SNR (e.g. 50, 100) examples. Notably, optimal performance was observed at intermediate SNR levels (e.g., SNR 25 and 35), highlighting a correlation between SNR spread and generalisation capabilities. Hence, to get the optimal performance we train these binary classifiers for each pair from mixed dataset i.e taking samples from all SNR values.

We first created a dataset with equal number of samples from different SNR values. We test the model performance by taking samples from SNR 15, 25, 35, 50 and 100 in ratios of 10\%, 15\%, 20\%, 25\%, and 30\%, respectively. We also tested the model performance by taking the ratio of samples in reverse order such that low SNR samples are slightly over-represented in the training set. We test the performance for these models separately for each SNR value. In most cases, we see better performance when we take equal or more samples from a lower SNR. We obtain good performance ($> 90$\% ) on most of the binary models except in the case of type IV vs II and type IV vs V. 
\par
To improve the performance for both these cases we used higher dropout rate upto $0.3$ compared to earlier values of $0.1$ and also input the frequency-averaged time series in parallel to the first dense layer. We see a trend in the case of type II vs type IV where more misclassifications happen for type II bursts having larger widths. With further tweaking in the architecture, we could get better accuracy in testing with different SNR values. Even after many changes to architecture in the case of type IV and type V the overfitting issue still remains. We discuss this specific case later in Section~\ref{sec:typeIVvsV}.
\par
We note that the observed SNR distribution of detected FRBs is strongly influenced by telescope sensitivity and selection effects, and therefore does not reflect the intrinsic population. Training on such a biased distribution can lead the model to learn instrument-specific features rather than robust burst characteristics. By using a balanced mixture across SNR values, the network learns noise-invariant features, improving generalisation and enabling more telescope-agnostic performance.

\subsection{Metrics}

 In our training, we have chosen accuracy as the primary metric for model evaluation since we do not have imbalanced datasets (by design).  To quantify the optimisation process, we employ the \texttt{binary\_crossentropy} loss function, which is minimised as the model trains. Similarly, For single multi-class classifier described in Section~\ref{sec:single_MC_classifier} we use \texttt{categorical\_crossentropy} as loss function.

Deep learning models typically comprise millions of trainable parameters, which can make them prone to overfitting if the training data volume does not match. To mitigate this issue, in addition to keeping the models small, various regularisation techniques are employed. One of the techniques we predominantly use is \texttt{dropout} layers. However, even with the application of these regularisation techniques, the model may still exhibit overfitting. A straightforward way to recognise this is when the training accuracy continues to improve, but the validation accuracy plateaus or starts to decrease.

To prevent overfitting, we implement an early stopping criterion, which monitors the validation accuracy. If, after a certain number of training epochs, the validation accuracy shows no improvement or starts to decline, the training process is halted to avoid further overfitting.   

\subsection{Type IV vs V}
\label{sec:typeIVvsV}

Among all the cases, we found that classifying between type IV and type V is the most challenging task due to the similarity in their morphology. Bursts with smaller drift rates of type V can be very similar to type IV bursts. The initial models we trained were overfitting for this case, and we attempted to tweak the parameters of the framework. However, when that did not work we combined the following approaches to improve the performance of the model:


\begin{figure*}
  \centering
  \includegraphics[ width=\textwidth]{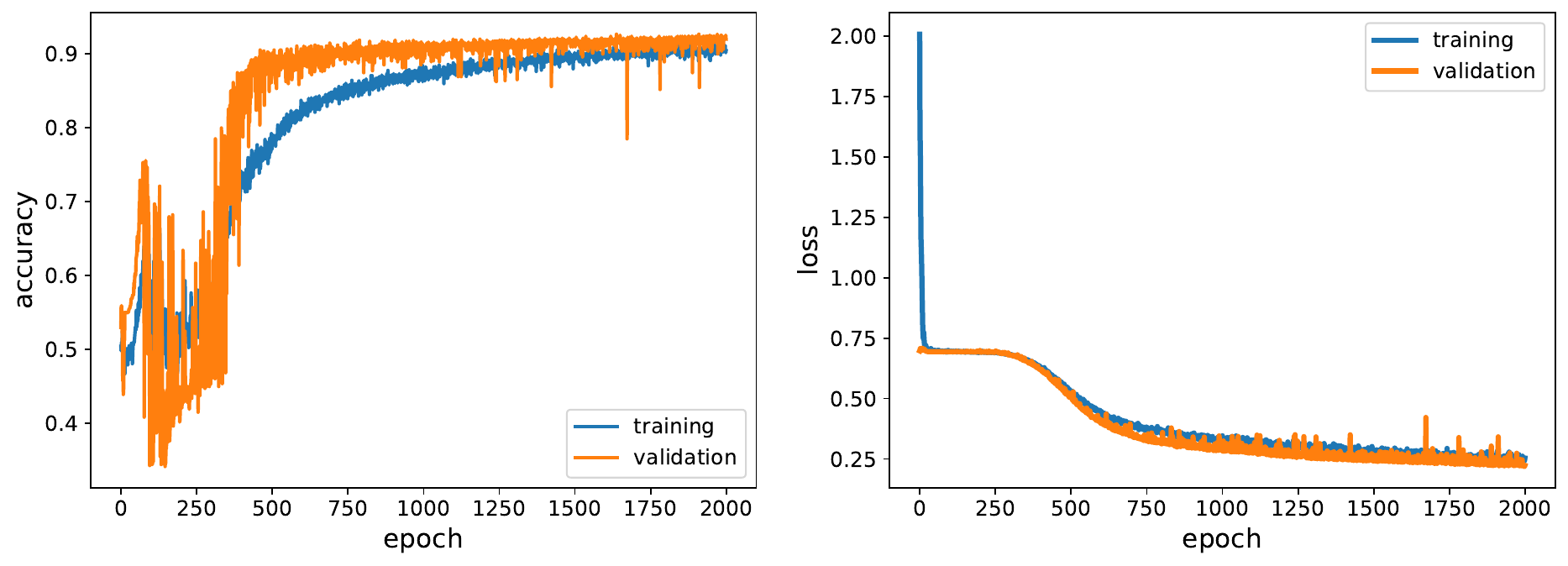}
  \caption{An instance of training a specific network with over 2000 epochs for binary classification of type IV vs type V.  \emph{Left Panel:} Accuracy vs. epoch for the training (blue) and validation set (orange). \emph{Right Panel:} Loss (binary cross-entropy loss function) as a function of epoch.}
  \label{fig:training_plots}
\end{figure*}

\begin{enumerate}
    \item Increasing the training set size, we simulate more samples for both types, which assist the deep learning model in learning those slightly unique patterns in both IV and V, which can aid in distinguishing both types.
    
    \item Increasing CNN layers and dense layer units: A smaller model may be insufficient to generate those distinct features for each type, especially the CNN layers responsible for extracting features in the image.
\end{enumerate}

In Figure~\ref{fig:training_plots}, we show an example where we trained a smaller model for a larger number of epochs where the model does not overfit.  After such tweaks we had well-performing models for all pairs of binary classifications. We then performed hyperparameter tuning on these base models to obtain optimal models as described next.

\begin{figure*}
  \centering
  \includegraphics[width=0.67\textwidth,scale= 0.9]{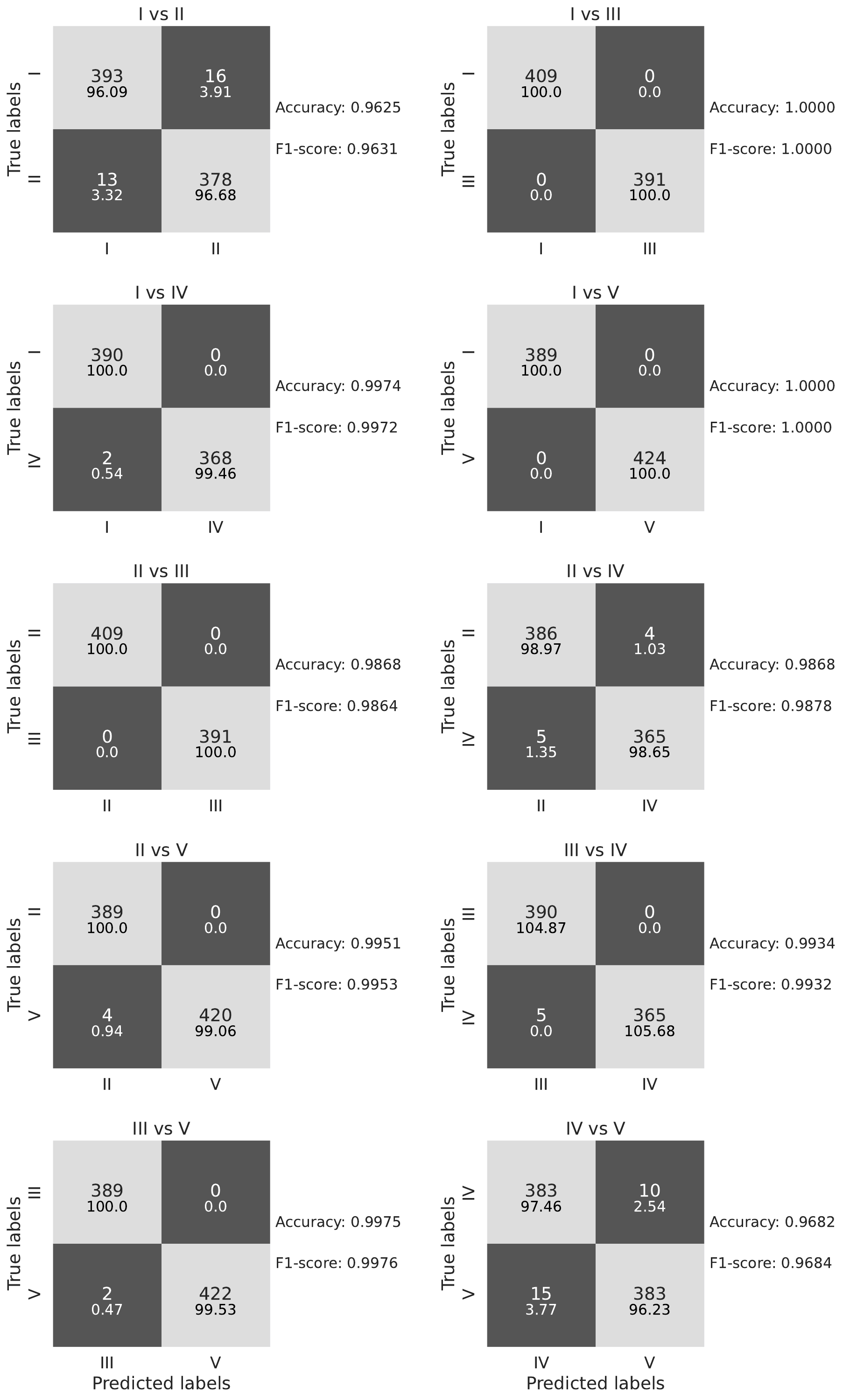}
  \caption{ We present pairwise binary classification confusion matrices. Each confusion matrix represents the inference on the test data (includes samples for all SNR values) from simulated dataset for an optimised model obtained by hyperparameter tuning for each of the binary classification. Light gray boxes represent the correct classifications and dark gray represent the incorrect classifications. Top number in each box represents the actual number and the bottom denotes the fraction of test samples for that particular type. Accuracy and F1-score for each case are shown on the right of each confusion matrix. }
  \label{fig:hyper_conf_mat}
\end{figure*}

\subsection{Hyperparameter Tuning}\label{sec:hyperparameter_tuning}
\label{sec:hyperparameter tuning}
 
Hyperparameter tuning, also known as hyperparameter optimisation, is a crucial step in optimising machine learning models by determining the most effective set of parameters that significantly impact learning and generalisation. These parameters, such as the number of filters in convolution layers, units in dense layers, dropout rates, and learning rates, are set before training and play a pivotal role in model performance.
Various methods, including grid search, random search, and Bayesian optimisation, can be employed for optimal hyperparameter search. In our approach, we utilised the \texttt{kerastuner} \citep{omalley2019kerastuner} with the \texttt{RandomSearch} method, conducting 100 trials to efficiently identify the best combinations of hyperparameters. We vary the hyperparameters such as number of nodes in the first dense layer and the second dense layer, learning rate, batch size, and dropout. Range of values taken for these hyperparameters described in Table~\ref{tab:Table_2}. The optimisation metric chosen for the evaluation was accuracy, with early stopping based on validation accuracy to prevent overfitting.

Upon obtaining optimised models, we further fine-tuned the binary classification by selecting an optimal threshold for decision-making. While a threshold of 0.5 is nominal, we used specific thresholds for each model to have the similar false positives and false negatives. In most cases since we see minimal misclassifications, we take threshold to be 0.5 for simplicity. For most models we see that there is an increase in the performance of the model compared to when trained with uniformly distributed widths. Accuracy of more than 95\% was achieved for all the models as shown in Figure~\ref{fig:hyper_conf_mat}. 

\begin{table}
\centering
\footnotesize
\caption{Parameters and ranges used for hyperparameter tuning.}
\label{tab:Table_2}
\setlength{\tabcolsep}{6pt} 
\begin{tabular}{lc}
\toprule
\textbf{Hyperparameter} & \textbf{Values} \\
\midrule
1st dense layer size   & 16, 32, 64, 128 \\
2nd dense layer size   & 8, 16, 32, 64 \\
Learning rate ($\times 10^{-5}$) & 1, 2, 5, 10, 20, 50 \\
Dropout fraction       & 0.15, 0.2, 0.25, 0.3, 0.35 \\
Batch size             & 32, 64, 128 \\
\botrule
\end{tabular}
\end{table}


\begin{figure}
  \centering
  \includegraphics[width=\columnwidth]{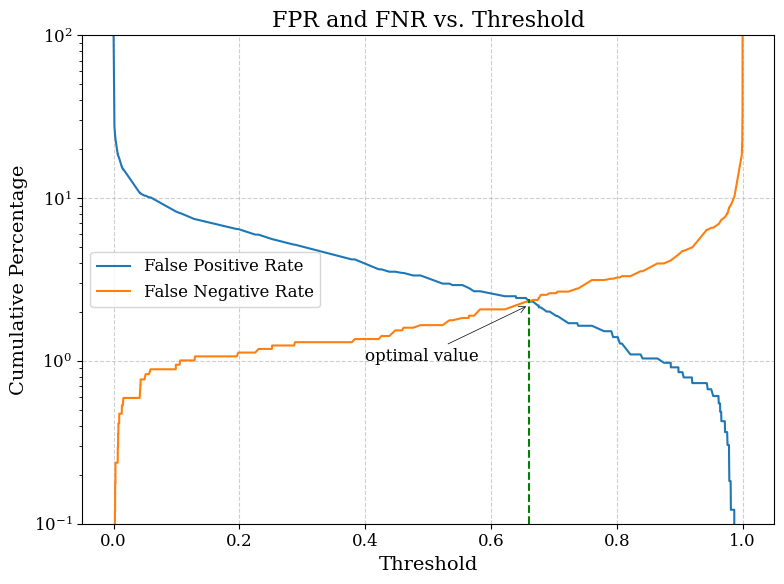}
  \caption{ False positive rate (FPR) and false negative rate (FNR) as a function of confidence output of the classifier. The intersection of the FPR and FNR curves signifies the value of confidence at which false positives equal false negatives. This specific example is for a type I vs II classifier. }
  \label{fig:optimal_threshold}
\end{figure}

For each set of hyperparameters, we use the average accuracy of two trials as the performance metric. We do not discuss about the models that include type VI since there are very few real examples among published FRBs. We determine optimal threshold for these 10 best models as illustrated in Figure~\ref{fig:optimal_threshold}. We use these models and thresholds to make a framework for classification with a set of binary classifiers which we describe in the next section.   

\subsection{ Multi-class classification with binary models}  \label{sec:multiclass_with_binarymodels}
\label{Multi-classwithbinary}
\begin{figure}
  \centering
  \includegraphics[width=\columnwidth]{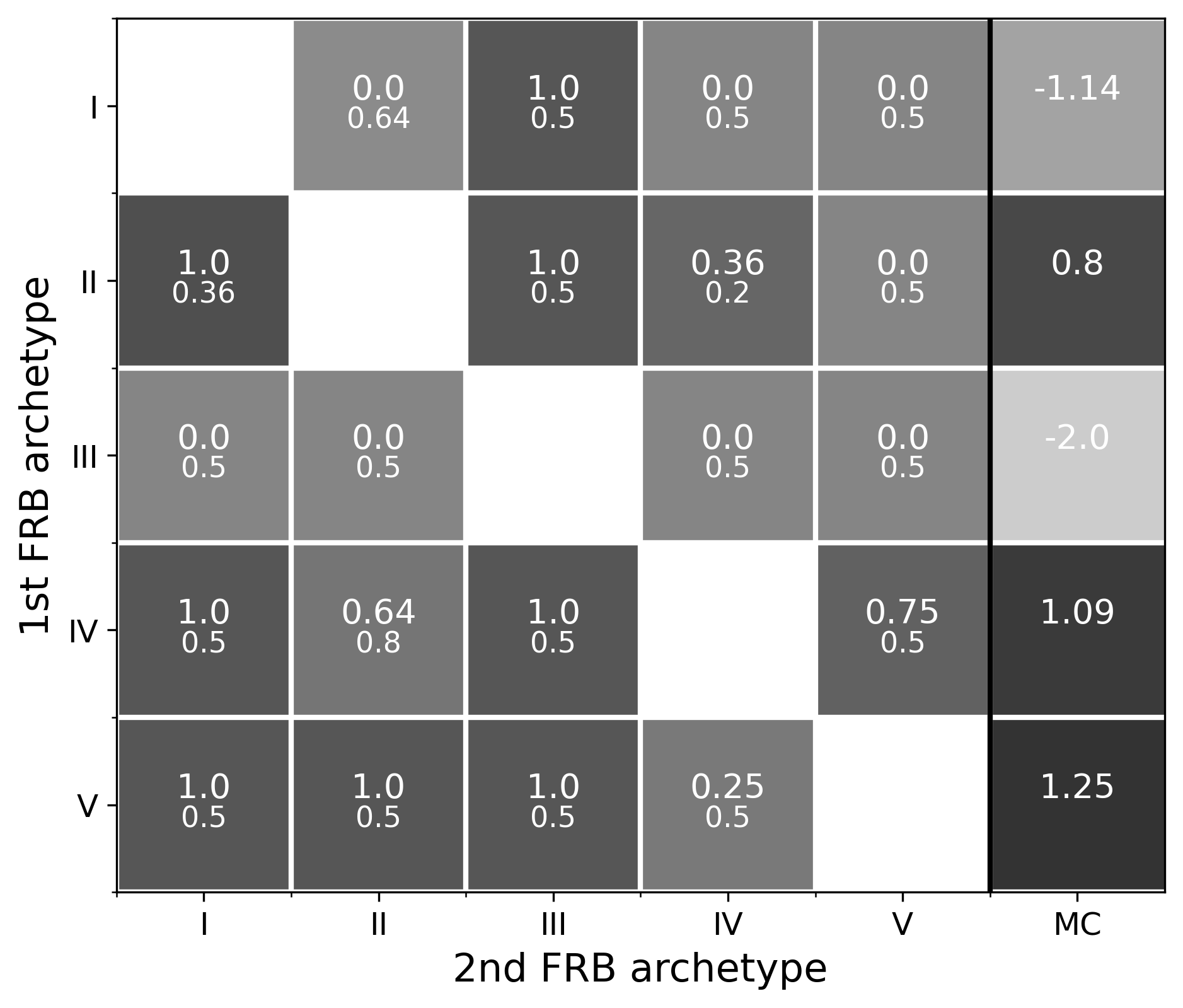}
  \caption{Example classification matrix for a single burst showing the output confidence from all the optimised binary models. Each element indicates the output score from a binary classifier corresponding to the archetype denoted by the row and column. Each element's upper value corresponds to the confidence i.e. the output confidence of the classifier while the lower value represents the optimal thresholds determined for that particular binary classification. Diagonal elements do not have any information. The last column is the sum of confidences of the that row after subtracted from the optimal threshold (i.e. augmented confidence). Here, a type V burst is classified through the multiclass classification matrix. It beats type IV narrowly, and type II is not far behind.}
  \label{fig:comb_matrix_sim_data}
\end{figure}

In this section, we outline our methodology for combining the confidence outputs from optimised binary classification models to perform classification for multiple classes\footnote{This is often referred to as \texttt{OnevsOne} classification in the machine learning domain.}. With optimised models and their respective thresholds for each binary case, we can construct an $N\times N$ matrix (where $N$ is the number of classes) for a given FRB, putting it through all binary classifiers. Figure~\ref{fig:comb_matrix_sim_data} shows an example of classification values. Each matrix element is the confidence output from the corresponding binary classifier, with complementary confidence in the conjugate element where positive and negative classes are interchanged. Diagonal elements, representing comparisons with the same class, are left out. The lower values in each element denote the  threshold for that specific classifier.

Once we obtain augmented confidence i.e confidence subtracted from the optimal threshold for each element of the matrix we can sum elements along each row. As an example, if we pass type V bursts to all the binary classifiers, it is expected that the binary classifiers involving type V bursts would have a higher value of output confidence than the rest of binary classifications. This means the row of type V bursts should have a greater value than the rest of the row for other types. We can now generate another column whose each element will represent the sum of augmented confidence for each row. We can classify this burst as type V correctly if the element corresponding to this type V in this column has the maximum value as illustrated in Figure~\ref{fig:comb_matrix_sim_data}. This framework serves as the foundation for classifying each of the five types: I, II, III, IV, V. Type VI is excluded from this analysis due to insufficient samples in the real test data.  
\subsection{ Testing with simulated data } 

\begin{figure*}{}
 \centering
 \includegraphics[width=0.8\textwidth]{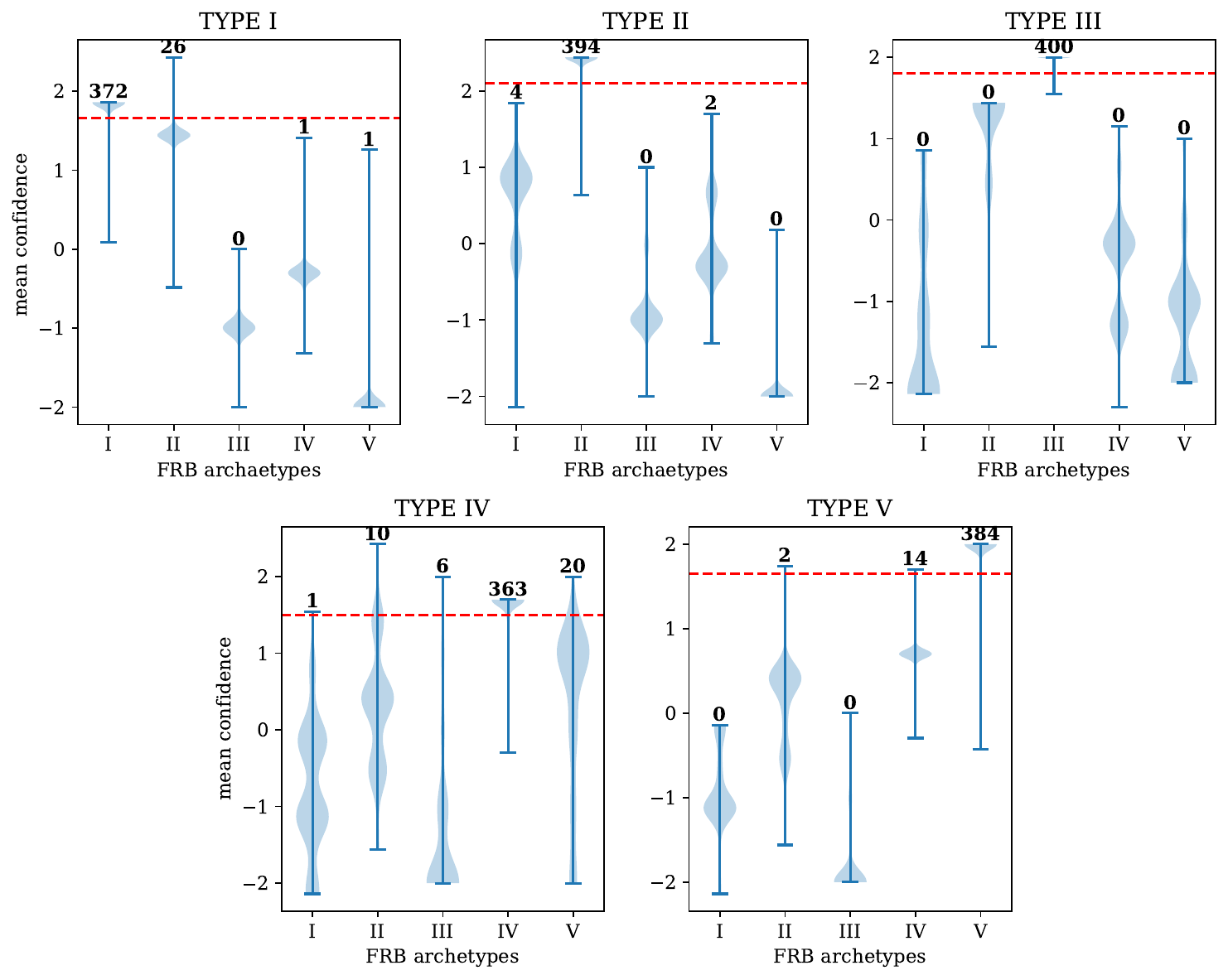}
 \caption{ Each violin plot in this figure displays distribution of the augmented confidences sum for one type as described in Figure~\ref{fig:comb_matrix_sim_data}. Each violin plot represents a total of 400 burst samples with a mix of SNR values. The red line indicates the approximate threshold where the augmented confidences sum is greater for the correct type compared to the other types. The number above each violin plot is the number of times the maximum value occurs for that particular type in the last column ouFigure~\ref{fig:comb_matrix_sim_data}. }
 \label{fig:violin_plots}
\end{figure*}

To test how well we can distinguish the classes with the framework we described earlier for multi-class classification we took 400 test samples of each FRB archetype including samples from all SNR values. We obtained the confusion matrix as shown in Figure~\ref{fig:comb_matrix_sim_data} for all these test samples. The distribution of each element in the last column in Figure~\ref{fig:comb_matrix_sim_data}, which corresponds to that particular type, is shown in Figure~\ref{fig:violin_plots}.  
We infer from these results that this framework works very well for classification with the simulated dataset. We can now test this framework on real data as well. For a better understanding, we plot the violin plot to show the distribution of the augmented confidence sum in these plots for each type. The violin plot allows a visual comparison of the distributions of the augmented confidence sum across the five types, which can be used to identify where the most misclassifications occur. 

\section{Testing with real data}
\label{sec:testing_with_real data}

We test the performance of the classifier with publicly available data from 535 FRBs bursts from the first CHIME/FRB catalog which is currently the largest homogenous dataset of FRBs currently available.

We first need to normalise and reshape the bursts to $256 \times 256 $ size for inputting into the classifiers. When we simulate the bursts we do not take into account the fact that every radio telescope will be affected by a dynamic RFI environment. The presence of narrow-band RFI leads to the masking of many frequency channels. Similarly, for CHIME, several channels are masked. On average, $\sim$30--40\% of channels are flagged for CHIME data but the frequency channels that are masked are largely random. The classifier has not been trained to handle these masked frequency channels. These masked channels are mostly random due to the dynamic RFI environment at the telescope site. It is important to note that several features in the dynamic spectra of FRB can be lost due to masking. We interpolate the data in the missing channels in order to make it compatible with our classifier. Another way to handle missing data is to train the classifier on data with randomly masked channels so the neural network learns to avoid missing information. There are several caveats to this approach and will be a part of future work.   

\subsection{Interpolating masked channels}
\label{sec:interpolation}

The publicly available data from the first CHIME/FRB catalog has the de-dispersed dynamic spectra and model data for each burst. The waterfall data has 16384 frequency channels and an adequate number of time samples to represent the burst. Each time sample has an integration time of $0.98304$ ms. We bin the frequency channels from 16384 channels to 1024 channels as interpolation on such a large number of channels is not useful.  

\begin{figure*}
 \centering
 \includegraphics[width=\textwidth]{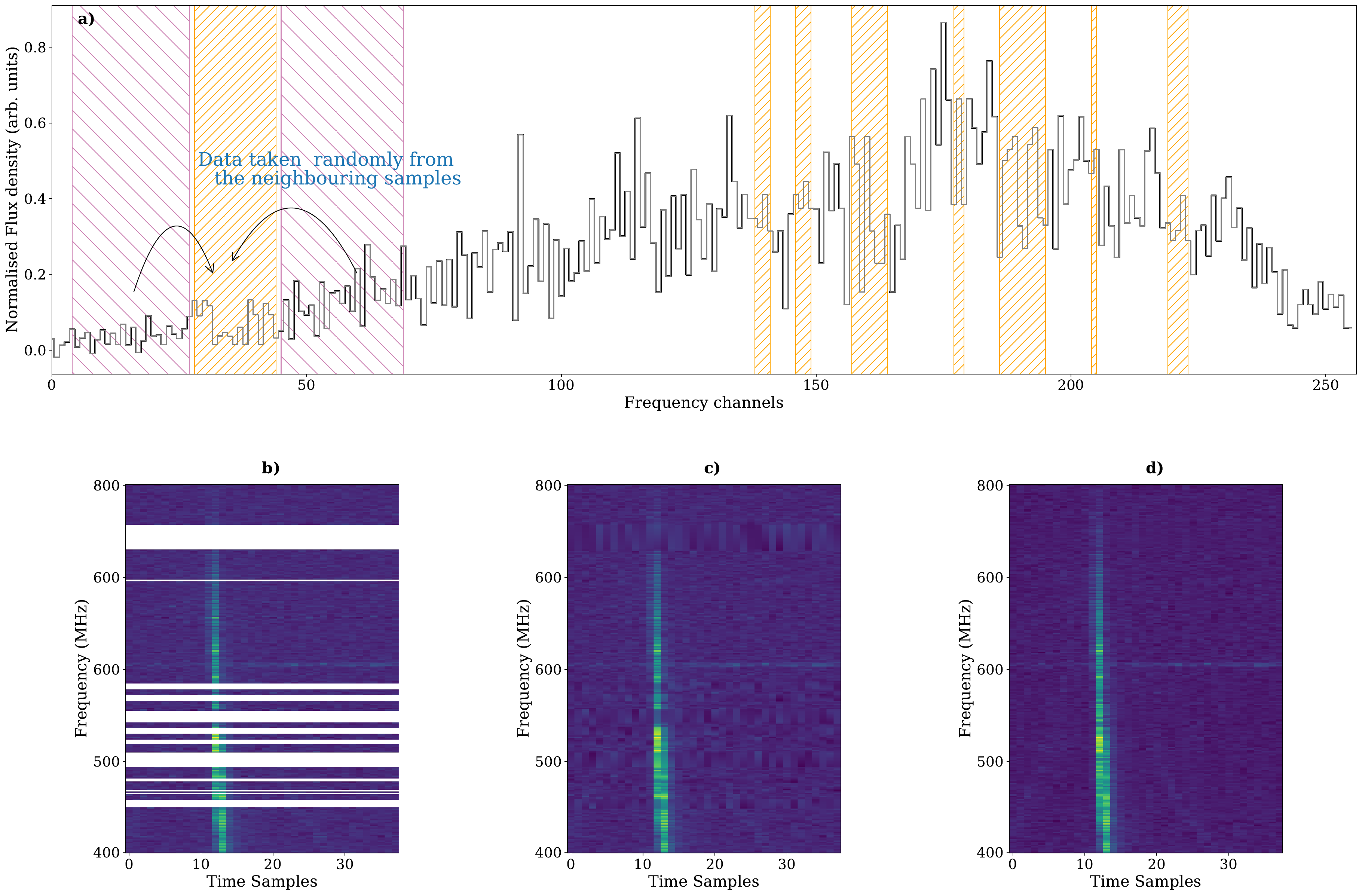}
 \caption{a) This figure shows the Intensity as a function of frequency channels for one particular time sample corresponding to FRB emission seen in dynamic spectra of one of the CHIME/FRB catalog. The masked channels are indicated by orange hashes. The pink hashes indicate the neighbouring region used to fill the masked channels. b) Dynamic spectra of one of the CHIME/FRB burst c) Dynamic spectra after using linear interpolation to fill the masked channels d) Dynamic spectra after interpolating as described in Section~\ref{sec:interpolation} and illustrated in panel a). }
 \label{fig:interpolation}
\end{figure*}

We interpolate the data independently for each time sample. Figure~\ref{fig:interpolation}, panel (a), shows an example of spectrum at a single time sample. We identify gaps of consecutive missing channels (yellow hatched regions in panel a \& whitespace in panel b) and identify a range of valid data from 1.5$\times$ the number of missing channels on either side of the gap (hatched pink region in panel a). We sampled (with replacement) the missing intensity data from the values of the valid neighbouring data. To avoid adding outliers, we eliminated any values in the valid neighbouring data that were outside of one median absolute deviation from the median of the valid neighbouring data.  

This is a more robust approach while it retains the morphological structure of the burst and also retains local noise properties. For comparison, Figure~\ref{fig:interpolation}, panel (c) shows standard linear interpolation applied to the waterfall plot in panel (b). The interpolation leads to structured noise and artefacts that are produced for time samples which do not have any detectable emission or have large gaps. Figure~\ref{fig:interpolation}, panel (d) shows the dynamic spectra generated from panel (b) using the method described above.


After the interpolation for the dynamic spectra is done we also pad the data in time to make the dynamic spectra of standard size ($256 \times 256$ pixels). We pad the data by adding Gaussian noise by matching mean and standard deviation to that of the local noise.    

\subsection{Test with the CHIME/FRB catalog}
\label{sec:test_with_chime_data}
\begin{figure}
 \includegraphics[trim={0.1cm, 0.1cm, 0.1cm, 0.1cm}, clip=false, width=\columnwidth]{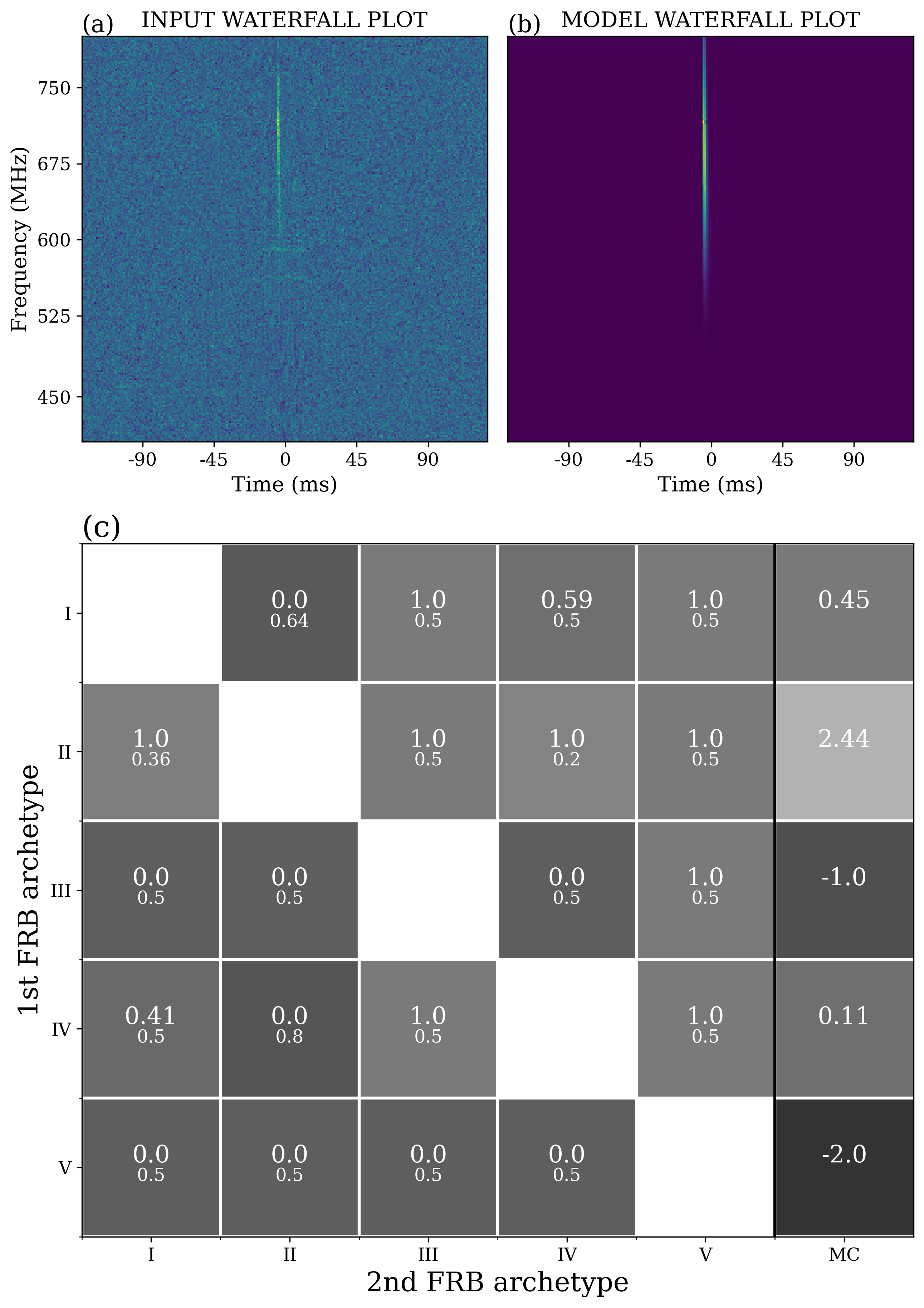}
 \caption{ a) Waterfall plot of one of the type II CHIME/FRB catalog bursts. b) Model waterfall plot for the same burst shown in panel a). c) This matrix is as described in Figure~\ref{fig:comb_matrix_sim_data} for panel a) as the input. The last column indicates the consensus class (type II in this case). }
 \label{fig:example_for_CHIME_burst}
\end{figure}

The CHIME/FRB Catalog \citep{2021ApJ...923....1P} consists of dedispersed waterfall plots as well as morphological model fits and fit parameters for each FRB. Applying our criteria from Table~\ref{tab:Table_1} and Section~\ref{sec:simulation} to the model fits, we identify 163, 270, 62, 20, and 20 bursts of types I, II, III, IV, and V respectively. There are no type VI bursts. Prior to classification, we reshape, interpolate and normalise the data to make it compatible with the training set. In Figure~\ref{fig:example_for_CHIME_burst}, one such example shows correct classification using a framework with a set of binary classifiers for a type II CHIME burst. 

We initially used a uniformly sampled width distribution to generate our training set, but the performance was poor. Upon closer inspection of the misclassifications, a recurring pattern emerges: many samples of types I and II are incorrectly classified as type IV, and vice versa. The misclassified samples indicate that a significant number of type I and type II bursts with larger widths are consistently misclassified as type IV and V.

\begin{figure}
  \centering
  \includegraphics[width=\columnwidth]{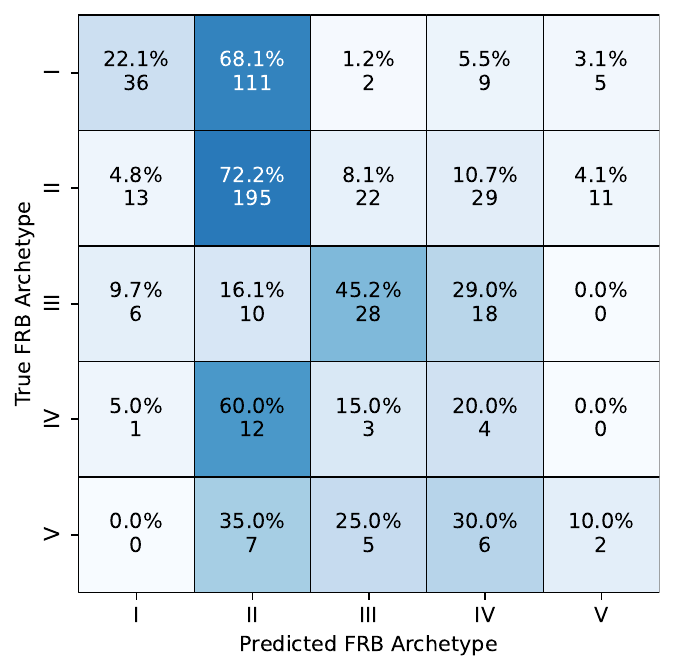}
  \caption{Confusion matrix after classifying the CHIME/FRB first catalog using the multi-class framework described in Section~\ref{sec:multiclass_with_binarymodels}. In each square, the upper number corresponds to the bursts classified in that combination of true (row) and predicted (column) archetypes. The lower number is the same as a fraction of the total number of bursts of that type in the CHIME/FRB catalog --- 163, 270, 62, 20, 20 samples for types I, II, III, IV, and V respectively.}
  \label{fig:test1_chime_data}
\end{figure}

This led us to use the width distribution described in Section~\ref{sec:simulation} for generating the training sets. With a more realistic width distribution, the classification was substantially improved. Using the optimised models described in Section~\ref{sec:hyperparameter_tuning}, we perform the classification for multiple classes. The overall classification results are shown in Figure~\ref{fig:test1_chime_data}. The overall efficiency still hovers around roughly 50\% for multi-class classification using the binary (\texttt{OnevsOne}) models. Since we could not achieve a good performance with five classes we also explored the 4-way and 3-way classifications (i.e. with reduced complexity). We present the results and discuss them in Appendix~\ref{sec:4wayclassification}.

 

\subsection{ Single Classifier }
\label{sec:single_MC_classifier}

\begin{figure}{}
    \includegraphics[width=\columnwidth]{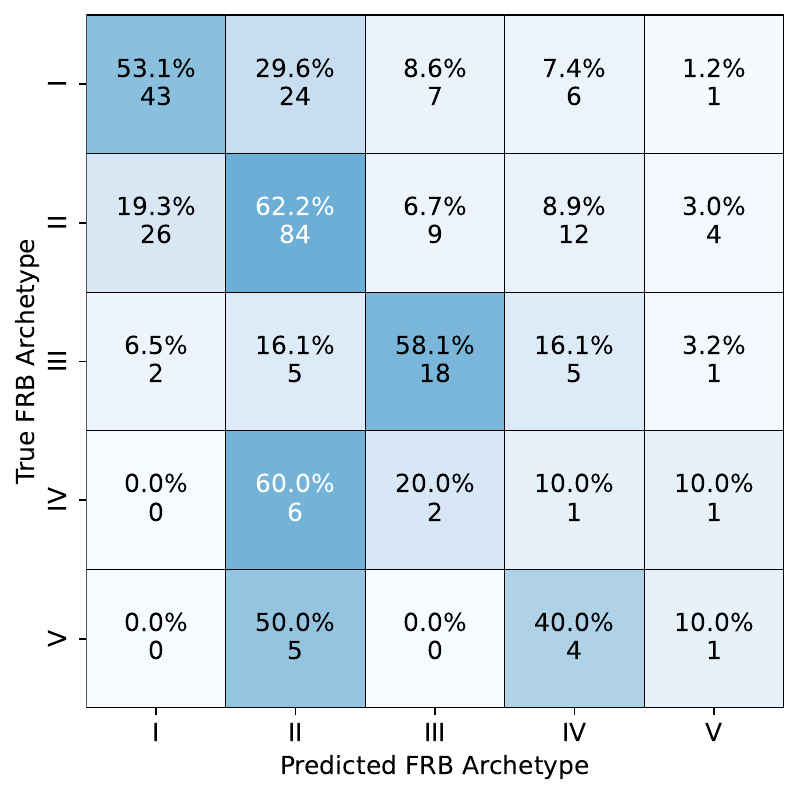}
    \caption{Confusion matrix for multi class classification using a single multi-class classifier. The details of the plot are the same as in Figure~\ref{fig:test1_chime_data} except that each type has half the number of bursts present in CHIME/FRB catalog. }
    \label{fig:test_with_single_model_multi}
\end{figure}

We pursued an alternative method by constructing a unified multi-class classifier i.e single network architecture. We designed a comparable but more extensive network illustrated in Figure~\ref{fig:architecture} for the \texttt{OnevsOne} classifiers, incorporating a \texttt{softmax} layer with five classes at the end. We employ a larger network with one more convolution layer and units in the dense layers. The training dataset described in Section~\ref{sec:simulation} was utilised, along with half of the CHIME/FRB catalog bursts. 
We roughly use 1800 simulated bursts including all SNR values for each archetype so in total we have 9000 samples for training. We also include half the number of bursts for each type in the first CHIME/FRB catalog and then the dataset is split into 80\% for training and 20\% for validation dataset. The remaining half of the bursts for each type are used for testing the classifier. Based on the best performing base model we get the final tuned model by exploring a similar set of hyperparameters as described in Table~\ref{tab:Table_2}. The accuracy of the classifier when tested with the simulated test dataset is $\sim$73\%. We test performance of the model with the remaining half of the CHIME/FRB catalog bursts for each type, we achieve an overall accuracy of 55\% and the classification metrics are shown in Figure~\ref{fig:test_with_single_model_multi}. In Figure~\ref{fig:test_with_single_model_multi} shows we are achieving an efficiency of $\sim62\%$ for type II compared to $\sim72\%$ with classification with binary models. On the contrary, the accuracy for type I is better compared to classification with binary models. As also discussed in Section~\ref{sec:testing_with_real data} since the data for type III, IV and V are limited, we cannot say much about the accuracy or the performance of the model in real data for these archetype. This will be explained better in future work, where we have enough samples for these types. Due to the poor performance for these types we explored the classification scheme if we reduce one or two archetype which is discussed later in Appendix~\ref{sec:4wayclassification}. We also discuss the explainability of the model predictions using SHAP in Appendix~\ref{sec:shap}. 

\section{Discussion} \label{sec:discussion}

Our framework is a step toward automating the categorisation of Fast Radio Bursts (FRBs) based on their morphological features. A real-time predictive model can play a crucial role in prioritising noteworthy events that could substantially contribute to our understanding of FRB sources and emissions. As datasets of FRBs grow, such a classifier could serve as a standard tool for statistical analyses, shedding light on key properties of the FRB population. 

To train the classifier, we employed simulated datasets due to the limitations and imbalances in the currently published set of FRBs. While we achieved excellent performance in multi-class classification on the simulated dataset, our results on real data, specifically the first CHIME/FRB catalog, fell short of our expectations. We acknowledge several factors contributing to this discrepancy, which will be explored in detail in subsequent work.

One such factor is the assumption of a Gaussian distribution for noise, a simplification that may not accurately mirror the complexities of real-world scenarios. We recognise the importance of training the classifier in the presence of complex noise specific to a particular telescope. This approach aims to evaluate the relative performance of the classifier under real observational conditions. Subsequently, we plan to develop a framework for retraining our architecture with real data from specific telescopes, anticipating significant improvements in classification accuracy.

Additionally, the impact of the bandshape (mostly the telescope beam or the signal chain) on the data needs careful consideration, especially regarding whether the real-time pipeline provides band-corrected waterfall plots. This aspect is crucial for a comprehensive understanding of the data and may warrant adjustments in our approach. For actual use, this framework will have to be re-trained with data from each telescope, incorporating the biases and systematics of the telescope data. We expect the performance to improve compared to the current behaviour where the simulated data may not be completely representative of the CHIME/FRB catalog.
The imbalanced nature of the test dataset further complicates the analysis, making it challenging to discern trends in misclassifications. Notably, types IV and V have limited samples in the CHIME/FRB catalog, making it uncertain whether the classifier's overall performance is sub-optimal for these categories because of structural reasons or due to the lack of proper training data. 
We have reconstructed the dynamic spectra of FRBs via a new interpolation method. While this method seems to visually achieve the true morphology of the FRBs, it may be better to explore masked CNNs \citep{2022arXiv220509616Y, CHEN2023110881} following the principle of masked image modelling \citep[MIM; e.g. ][]{2021arXiv210608254B, 2022arXiv221010615P}. By introducing frequency-channel-wise masking as a custom image augmentation layer (similar e.g. to \texttt{keras RandomCrop}), we can train the classifier to handle the random masked channels. This would help the classifier better identify burst morphologies in the presence of masked narrow-band RFI in real data without the need for interpolation.
In addressing these challenges, we propose the exploration of alternative neural network architectures that excel in feature extraction. Transfer learning offers a potential solution, involving the use of pre-trained models for feature extraction, followed by training only the dense layers and the last layer of the Convolutional Neural Network (CNN). This approach could enhance the model's ability to generalise across different classes.
Looking ahead, our future work involves testing the framework on larger datasets with sufficient samples for each FRB type. We aim to refine the architecture to achieve optimal performance. These steps are integral to advancing the reliability and effectiveness of our FRB classification model. 
While this work focuses on the supervised classification of known FRB morphologies (types I–V), an important goal is the identification of anomalous bursts that do not belong to any predefined class. In practice, such anomalies cannot be treated as an additional labelled class, as their morphologies are not known a priori. Instead, anomaly detection can be naturally implemented using deep learning models trained only on common or well-characterised burst morphologies, and flagging events that are inconsistent with the learned representation. For example, convolutional autoencoders or variational autoencoders trained to reconstruct typical FRB dynamic spectra can identify anomalous bursts via large reconstruction errors (e.g. \citep{Villar_2021}). Alternatively, outlier detection in the latent feature space of a trained classifier can flag anomalous cases using methods such as isolation forests or distance-based metrics. These methods have been shown to be effective in identifying rare or novel transients in large astronomical datasets \citep{Lochner_2021,Anderson_2025,chaini2025searchunknownunknownsmultimetric,Crispim_romao_2025,gupta_2024_anamoly}. Such approaches would allow the existing framework to flag potentially interesting FRBs without requiring explicit training on anomalous classes, and provide a scalable path toward automated follow-up prioritisation.   
\subsection{Future Extensions}
Here, we discuss possible future extensions (in increasing order of challenge) of this classification framework and how we can incorporate more features in our classification of FRBs.
\subsubsection{Classes identified from data} 
Currently, we define classes by visual inspection of the bursts. We can leverage dimensionality reduction algorithms to identify different classes in the FRB dynamic spectra and simulate those to understand if they are particularly related to different types of progenitors. \citet{2023ApJ...945...67S} used \texttt{UMAP} and \texttt{t-SNE} to produce embeddings for multi-band GRB data which very cleanly classify short and long GRBs, independent of the traditional classification features, $T_{90}$ and hardness ratios. \citet{2023MNRAS.522.4342Y} applied similar concepts to the dynamic spectra of CHIME/FRB bursts spectrograms to differentiate repeaters from apparent non-repeaters. Their clustering algorithms finds about six specific clusters within the repeaters and non-repeaters. \citet{2023arXiv230701054M} identifies anomalies as well as typical classes of the RFI instances based on auto-correlation of spectrograms through novel self-supervised learning for radio telescope health monitoring. We can extend our framework to detect anomalies -- for anomalous objects the output confidence of the  classifiers -- binary as well as multi-class -- will be away from typical/expected classification thresholds. By looking at FRBs that follow this trend for all classifiers we can identify the anomalies.
The sample size of known FRBs with high signal to noise ratio that can be used to train such methods to identify new classes of FRBs is limited. This pushes us to start with visually identified, ad-hoc classes of FRBs. However, in the future, we can use classes based on known FRBs to have a more physical classification. Another challenge with the physically identified classes is that it will not have a balanced representation of different FRB types. 

\subsubsection{Generative AI techniques}
Recently, Generative AI has found application in generating images that are close to real with text inputs \citep{Iangoodfellow2014}. Similar application can be useful in many areas for creating a more robust training dataset \citep{XinGANsreview2023}. Generative AI techniques need carefully set guardrails in order to be able to generalise the training set while preserving the essential features of the classes.  Here, it will be more useful to have adversarial networks - one part creating samples, and another ruling about their goodness, and in turn both learning to do their jobs better - the first one creating more and more realistic samples, the second one becoming better at discerning between samples of a given class and those not belonging to that type. As our binary and multi-class classifiers improve we also plan to put together such adversarial networks.

\subsubsection{Hierarchical classification} 
There are domain-specific questions about FRBs that we could ask in a hierarchical manner: e.g. is the FRB single component or multi-component? does it have a narrow-band spectrum? does a single spectrum describe all the components in the FRB? is there periodicity among the multiple components? These questions are physically motivated and are independent of the assigned classes. This would be somewhat akin to the approach in \texttt{GWSkyNet-Multi} \citep{abbott2022_hierarchicalclassification, 2023arXiv230812357R} for GW events from the LIGO-Virgo-Kagra collaboration. There a top-level binary classifier differentiates between merger events and astrophysical glitches. Then in the second layer, GW merger events are separated into black hole-black hole merger events and other merger events and so on. For FRBs, we envision that we can develop a similar scheme -- e.g. RFI vs FRB $\rightarrow$ single vs multiple components $\rightarrow$ single spectrum or downward drifting etc. will prove fruitful.

\subsubsection{High time resolution}  
We currently do classification for data sampled at ~1 ms resolution but recent studies show microstructure in bursts from repeating FRB 20180916B, and FRB 20121102A. The microstructure could be buried for bursts where we have coarser resolution \citep{2023MNRAS.526.2039H, Dayetal, 2023arXiv231214133F}. There is an increasing number of events that have  recently shown structure at microsecond timescales, especially for CHIME/FRB which detects a high number of FRBs everyday. We can extend the scope of classifier by training it with high time resolution data.


\subsubsection{Polarisation} 
The polarisation properties of FRBs have been found to be intriguing, with several bursts exhibiting diverse morphologies \citep{2021NatAs...5..594N}. Bursts with seemingly similar intensity profiles (say, type I) can have different linear or circular polarisation properties. Our classification is currently limited to Stokes I. However, expanding this framework to include all four Stokes components and classifying them could provide us with additional insights into the nature of these bursts. This effort is challenging due to the added complication of simulating a variety of polarisation behaviours, from flat position angle, linear polarisation \citep{2021NatAs...5..594N}, a rotating-vector-model-like position angle variation \citep{2020Natur.586..693L}, circular polarisation \citep{2022MNRAS.512.3400K, 2022Natur.609..685X}, and even more complex Faraday conversion among different sub-bursts \citep{2020ApJ...891L..38C}. With larger FRB sample sizes and with the help of generative AI, we may be able to generalise some of the polarisation properties of different FRB classes.

\section{Summary}
\label{sec:summary}

In this work, we demonstrate a working automated classifier that with further improvements and additions will help in following up interesting and anomalous morphologies of FRB events that can be crucial to understanding FRB origins and their properties. The classifier will also be useful in statistical studies of FRB morphology as the number of FRBs increase significantly. To develop such a classifier we identified the need to simulate FRBs due to limited number of published FRBs. We presented a framework for simulating six major types of FRB morphology based on the first CHIME/FRB catalog that will be made available for public usage in near future. We generate a training dataset for classifying between each pair of the identified FRB archetypes. We train these networks which are based on CNN architecture to classify each pair after which we obtain optimised models that are used to build a multi classification framework. We obtained excellent performance on the simulated dataset for the multi classification framework. The test with real data from first CHIME/FRB catalog bursts gives a sub-optimal performance for reasons we have have discussed earlier. In the future, real data will aid us in improving the performance of the classifiers. Our next work will address the issues in achieving optimal performance for the automated classifier which can be used for prompt followups for interesting, rare and bright FRB events.   

\section*{Acknowledgements}

A. K. would like to thank Sujay Mate for useful discussions. S.P.T. is a CIFAR Azrieli Global Scholar in the Gravity and Extreme Universe Program. This research was supported in part by a generous donation (from the Murty Trust) aimed at enabling advances in astrophysics through the use of machine learning. Murty Trust, an initiative of the Murty Foundation, is a not-for-profit organisation dedicated to the preservation and celebration of culture, science, and knowledge systems born out of India. The Murty Trust is headed by Mrs. Sudha Murty and Mr. Rohan Murty.

\section*{Data Availability}

We have released a public GitHub repo containing the code and data related to this manuscript. You can find the repo through this link https://github.com/ajay2609/frabjous/. 

\section*{Software}
Matplotlib \citep{Hunter:2007}, keras \citep{chollet2015keras}, tensorflow \citep{tensorflow2015-whitepaper}, scikit-learn \citep{pedregosa2011scikit}



\appendix
\label{sec:appendix}

\section{Prescription for simulating type VI bursts}\label{sec:typeVI_prescription}
We here provide the formulation we adopted to simulate type VI bursts. However, the examples are shown in Figure~\ref{fig:example_bursts}. 

\paragraph{Type VI:} The parameters are kept similar to the type I bursts, but we multiply the spectrum with the diffraction pattern of the telescope \citep{lin2023a} though we note that an FRB of any morphology can be detected in the far sidelobe. We model the spectrum $I(\nu, \theta)$ through Equation~\ref{eq:equation_diff} where the off-axis angle $\theta$ represents the position of the FRB in the far sidelobe away from the CHIME/FRB meridian, A is the intensity at the $t\theta$ equal to zero, $\nu$ is the frequency, and $a$ is the aperture of the telescope, and $c$ is the speed of light. This implementation is specific to CHIME/FRB, but it can easily be modified to mimic instrumental signatures of other telescopes.
\begin{equation}
     I(\nu, \theta) = A \left(\frac{\sin(\phi)}{\phi}\right)^2,
     \phi = \frac{2\pi a \nu}{c} \sin(\theta).
     \label{eq:equation_diff}
\end{equation}

\begin{enumerate} 
    \item We choose a off-axis angle $\theta$ uniformly between [2, 40] degrees which represents the position of the FRB in the far sidelobe away from the CHIME/FRB meridian.
    \item We have used an aperture size of $80$ m for CHIME.
    \item We generate an FRB as per the specifications for type I and then multiply the spectral response from Equation~\ref{eq:equation_diff} before renormalising the flux of the FRB.       
\end{enumerate}

\section{SNR comparison for boxcar and T90 method} \label{sec:snrcomparison}
To enable a comparison between the boxcar S/N which is widely used in FRB searches and the $T_{90}$ definition, we compile a set of single-component bursts from the first CHIME/FRB catalog. The simplicity of these bursts minimizes ambiguities in defining burst duration and S/N, allowing for a more direct comparison. Their key properties are listed in the Table~\ref{tab:frb_snr}.
\begin{table}[htbp]
\centering
\caption{SNR and width measurements for selected FRBs from CHIME/FRB bursts using boxcar and T90 definition}
\label{tab:frb_snr}
\begin{tabular}{lcccc}
\hline
FRB Name & Boxcar SNR & $t_{90}$ SNR & Boxcar Width (bins) & $t_{90}$ Width (bins) \\
\hline
FRB20180725A & 35.01 & 18.87 & 4 & 24 \\
FRB20180727A & 8.91 & 5.35 & 4 & 27 \\
FRB20180729A & 102.99 & 83.02 & 1 & 3 \\
FRB20180729B & 12.87 & 6.01 & 2 & 28 \\
FRB20180730A & 42.49 & 40.85 & 4 & 11 \\
FRB20180801A & 19.88 & 14.97 & 8 & 41 \\
FRB20180806A & 14.90 & 8.30 & 2 & 21 \\
FRB20180810A & 11.63 & 5.94 & 2 & 33 \\
FRB20180810B & 43.05 & 34.09 & 1 & 4 \\
FRB20180812A & 9.80 & 7.83 & 8 & 36 \\
\hline
\end{tabular}
\end{table}
\section{3-way \& 4-way classification}
\label{sec:4wayclassification}
For multi-class image classification, the complexity of the problem increases with number of classes. Typically, we require more complex and bigger models to extract specific features for each class as the number of classes increase. These classes are not physically motivated but are based on their distinct features in dynamic spectra. In certain cases one type can be very similar to other type which leads to confusion for the model. Here we explore and discuss the performance of models with fewer classes. 

We evaluate the performance of our framework with the CHIME/FRB dataset for 4-way classification instead of 5-way. For multi-class classification with binary models (\texttt{OneVsOne}), we do not need to change the architecture or retrain models but we simply use the binary classifiers relevant for our 4-way classification. For e.g., we exclude all models that are related to type I to do a 4-way classification. In this case, we see an improvement in the overall accuracy by 11\% compared to 5-way classification. In the case where we exclude type II, we see an improvement in the overall accuracy by 18\%. This also suggests that type II has more cases that are similar to type IV compared to type I. Many of the misclassifications for type IV are as type I and II. This indicates the distinct multi components feature for type IV is not picked by the models. These two inferences are illustrated by left and middle panels in Figure~\ref{fig:4-wayclassifcation_forbinarymodels}. In case excluding type II we see performance for the type I is 86\% and the overall accuracy is 68\%. While in the case where we exclude type I, the accuracy for the type II is 70\% and the overall accuracy is 61\%. With the 3-way classification the overall accuracy is 52\% although the accuracy for type III and type IV are 61\% and 75\%.

\begin{figure*}
\includegraphics[width=0.3\textwidth]{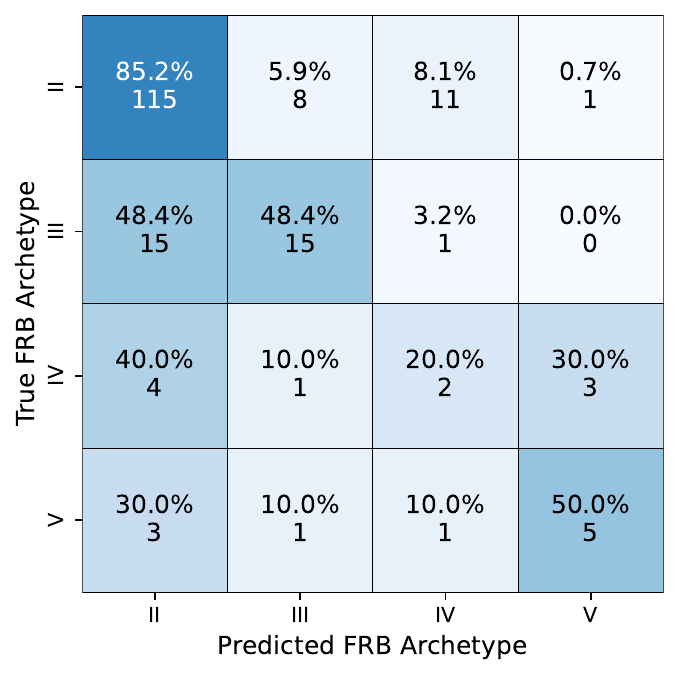}
 \hfill
\includegraphics[width=0.3\textwidth]{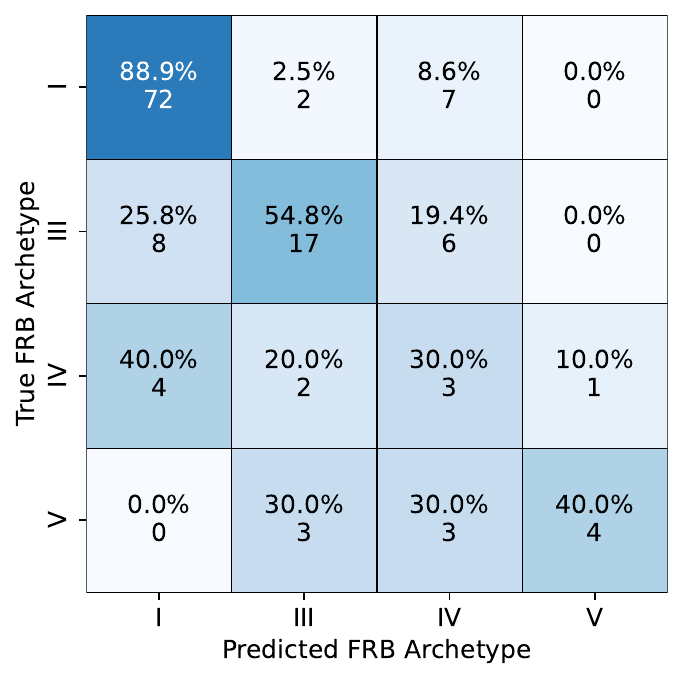}
 \hfill
\includegraphics[width=0.25\textwidth]{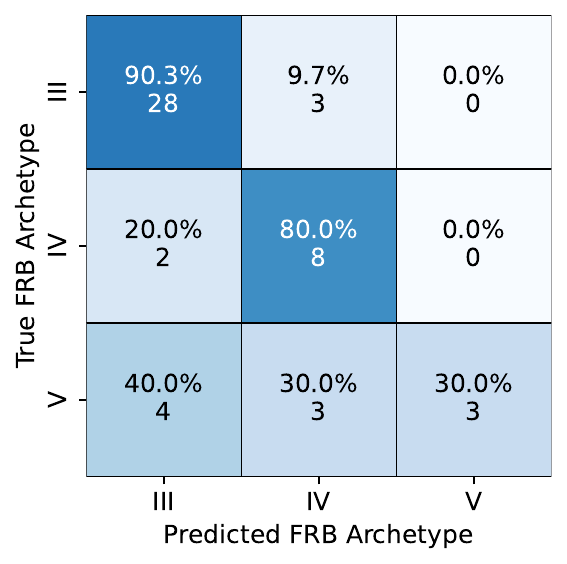}
 \caption{ \emph{Left Panel:} Confusion matrix with half of CHIME/FRB test catalog for the type II, III, IV, and V after we get optimised model as described in the text. In each element, the value above is the number corresponding to predicted archetype and actual archetype. The values below is the percentage of the particular type. \emph{Middle Panel:} Similar to the left panel, the confusion matrix for type I, III, IV, V. \emph{Right Panel:} Similar to left panel, the confusion matrix is for type III, IV, V. }
 \label{fig:4-wayclassifcationforsinglemodels}
\end{figure*}

\begin{figure*}
\includegraphics[width=0.3\textwidth]{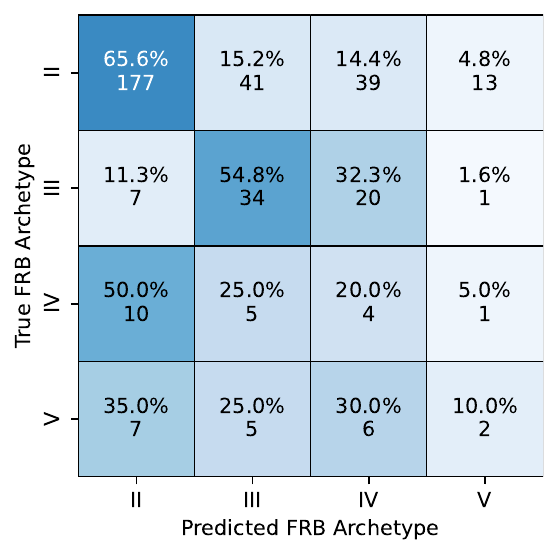} 
 \hfill
 \includegraphics[width=0.3\textwidth]{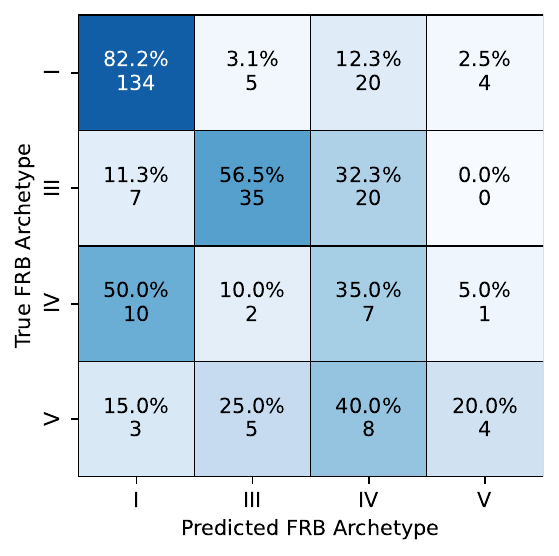} 
 \hfill
 \includegraphics[width=0.25\textwidth]{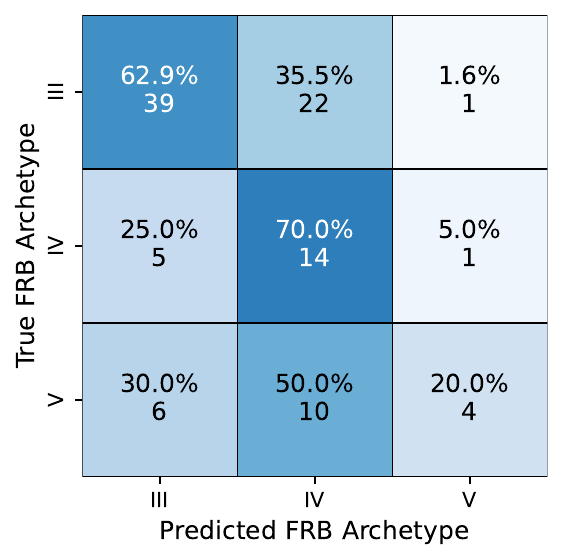} 
 \caption{ \emph{Left Panel:} Same as Figure~\ref{fig:4-wayclassifcationforsinglemodels} but with all the CHIME/FRB catalog bursts using the binary optimised models for each pair of classification. \emph{Middle Panel:} Similar to left panel, the confusion matrix for type I, III, IV, and V. \emph{Right Panel:} Similar to left panel, confusion matrix for type III, IV, and V.}
 \label{fig:4-wayclassifcation_forbinarymodels}

\end{figure*}

We also test the 4-way classification and 3-way classification using the single-multi classifier. We do a coarser optimisation with 5 iterations for the same set of hyperparameters and base model used in the single multi-class classifier as described in Section~\ref{sec:single_MC_classifier}. With retraining, we see a considerable increase in overall accuracy with 4-way classification and the results are shown in Figure~\ref{fig:4-wayclassifcationforsinglemodels}. In cases where we do classification excluding type I and type II, we see an overall accuracy of 74\% and 73\% respectively which is significant improvement compared to the 55\% overall accuracy with all five types.


\section{Model explainability with shap } 
\label{sec:shap}

\begin{figure*}
\centering
 \includegraphics[trim = 2cm 9.5cm 2cm 0.5cm ,clip,scale=0.50]{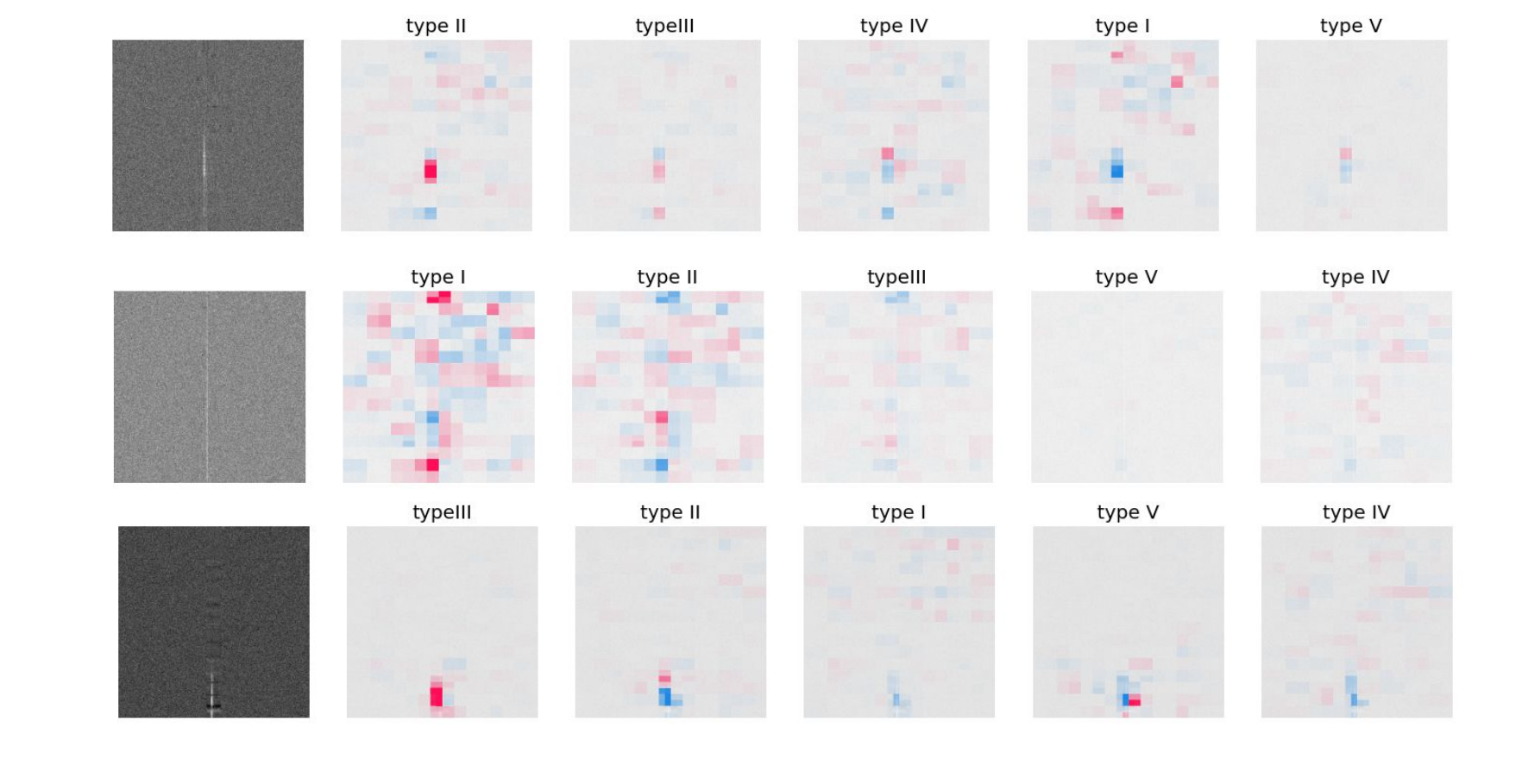}
 \caption{ SHAP values for three correctly classified burst examples from the CHIME/FRB catalogue. In each row, the left column shows the preprocessed input waterfall and the next five columns show SHAP value outputs of each classifier, ordered in decreasing order of classification confidence. In each SHAP plot, the pixels in red (blue) contribute positively (negatively) to the prediction of the model. From top to bottom, the input types are type II, I, and III, respectively. SHAP values indicate that feature extraction is good for types II and III. But for type I bursts, pixels areas corresponding to the background seem to contribute substantially to the classification. We generally see that SHAP values are higher (in red) at the end of the observing band on both sides or lower at one end (in blue) for type I when they are correctly classified. 
 }
 \label{fig:correct_shap_plots}
\end{figure*}

\begin{figure*}
\centering
 \includegraphics[trim = 0.6cm 12cm 2.5cm 1cm ,clip,scale=0.60]{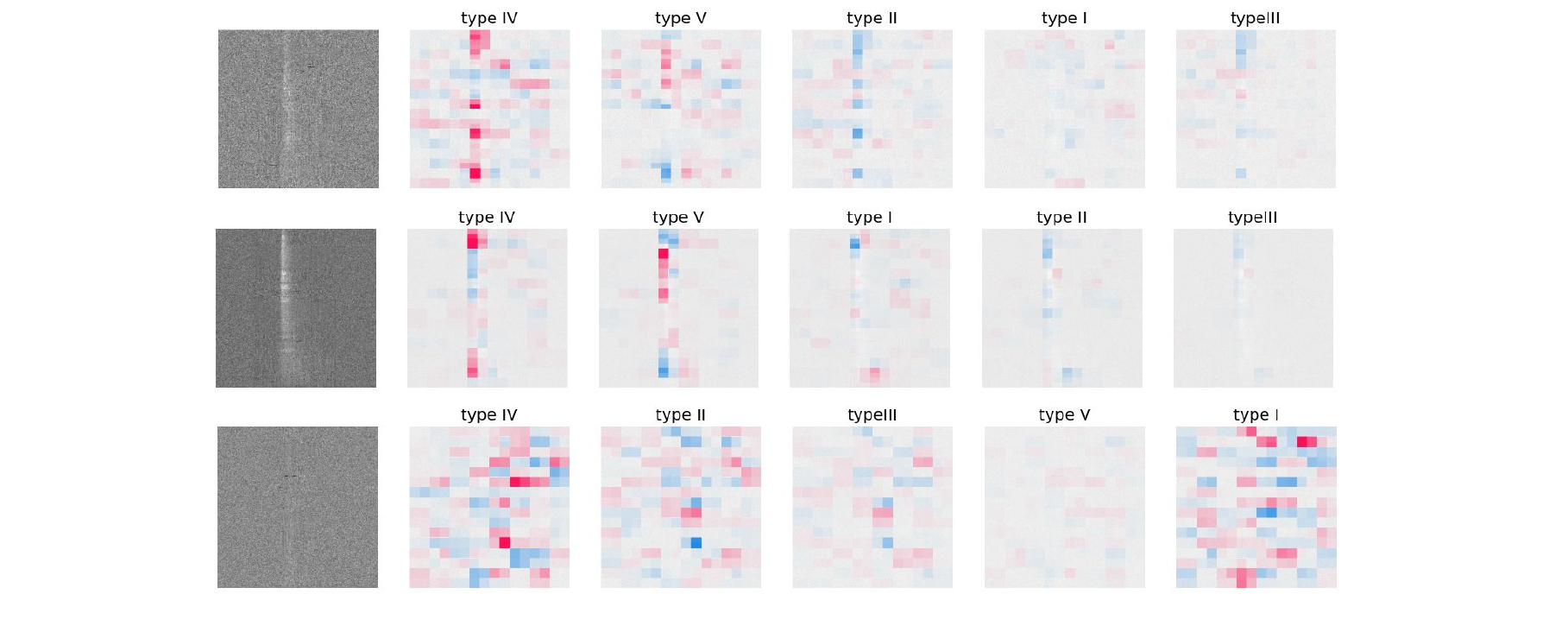} 
 \caption{ Same plots as Figure~\ref{fig:correct_shap_plots} but for incorrectly classified examples. From top to bottom, these are: 1) type I is misclassified as type IV. 2) type I misclassified as type IV due to scintillation seen in this case. 3) type II misclassified as type IV due to low S/N of the burst and fixating on noise.}
 \label{fig:shap_plots}
\end{figure*}


For interpreting complex predictions generated by deep learning models, SHAP (SHapley Additive exPlanations\footnote{\url{https://shap.readthedocs.io/en/latest/index.html}})\citep{NIPS2017_8a20a862} is a game theory based approach to explain the outcomes of machine learning models. The SHAP algorithm undertakes several trials with slight variations in the input parameters to measure the positive or negative contributions of each input (here, groups of neighboring pixels), to the final classification. 
We used the SHAP `explainer' specific to deep learning models with 5000 evaluations to understand the relative contributions of different parts of the dynamic spectrum to the final classification. This still produces coarse explanations but gives a better visualisation of the features that play an important role in the predicted outcome. We run the SHAP explainer using the single multi-class classifier (corresponding to the confusion matrix in Figure~\ref{fig:test_with_single_model_multi}) on half of the CHIME/FRB catalog bursts. Many of the misclassifications are due to low S/N of the bursts. In one the examples shown in Figure~\ref{fig:shap_plots},  the dynamic spectrum has artefacts even after interpolation which affects the model prediction. This happens mostly if there are many masked channels nearby. In Figure~\ref{fig:correct_shap_plots}, we see that band-limited feature is extracted properly in SHAP plots for type I and type II bursts.

\bibliography{filtered1}
\bibliographystyle{apj}

\end{document}